\documentclass[11pt,a4paper]{article}
\usepackage[utf8]{inputenc}

\def\papertitle{\bf Inverse Problem Approach for Non-Perturbative QCD: \\ Applications} 
\usepackage{authblk}
\title{\papertitle}
\author[1]{Yong Zheng\footnote{Email: yzheng2018@lzu.edu.cn}}
\author[1]{Ao-Sheng Xiong\footnote{Email: xiongash21@lzu.edu.cn, corresponding author}}
\author[1]{Fu-Sheng Yu\footnote{Email: yufsh@lzu.edu.cn, corresponding author}}
\author[2]{Ting Wei\footnote{Email: tingwei@lzu.edu.cn}}
\affil[1]{Frontiers Science Center for Rare Isotopes, and School of Nuclear Science and Technology, Lanzhou University, Lanzhou 730000, China}
\affil[2]{School of Mathematics and Statistics, Lanzhou University, Lanzhou 730000, China}
\usepackage[top=1in, bottom=1.25in, left=1in, right=1in]{geometry}
\usepackage{microtype}
\usepackage{lineno}
\usepackage{xspace}
\usepackage{booktabs}
\usepackage{comment}

\usepackage{caption} 

\usepackage{graphicx}  
\usepackage{color}
\usepackage{colortbl}
\DeclareGraphicsExtensions{.pdf,.PDF,png,.PNG}

\usepackage{amsmath} 
\usepackage{amsfonts,amssymb,amsthm,bm,braket,yhmath}
\usepackage{upgreek} 

\usepackage{hyperref}
\usepackage{hyperxmp}
\usepackage{hypcap} 
\hypersetup{hidelinks} 
\usepackage{txfonts}
\usepackage{amssymb}
\usepackage{cases}
\usepackage{algorithm}
\usepackage{algpseudocode}
\usepackage{mathrsfs}
\usepackage{color}
\usepackage{amscd,amssymb,amsthm,amsmath,bm,graphicx,psfrag}
\usepackage{epsfig,mathrsfs}
\usepackage[bf,SL,BF]{subfigure}
\usepackage{amsmath}
\usepackage{pict2e,keyval,fp}
\usepackage[bf,SL,BF]{subfigure}

\theoremstyle{definition}

\numberwithin{equation}{section}

\def\kaon   {{\ensuremath{K}}\xspace}
\def\Kbar    {{\kern 0.2em\overline{\kern -0.2em \kaon}{}}\xspace}

\def\KorKbar    {\kern 0.18em\optbar{\kern -0.18em K}{}\xspace}

\def\Dbar    {{\kern 0.2em\overline{\kern -0.2em D}{}}\xspace}

\def\DorDbar    {\kern 0.18em\optbar{\kern -0.18em D}{}\xspace}

\def\Lbar        {{\ensuremath{\kern 0.1em\overline{\kern -0.1em\Lambda}}}\xspace}
\def\LorLbar    {\kern 0.18em\optbar{\kern -0.18em \PLambda}{}\xspace}

\newcommand{\tev}{\ensuremath{\mathrm{\,Te\kern -0.1em V}}\xspace}
\newcommand{\gev}{\ensuremath{\mathrm{\,Ge\kern -0.1em V}}\xspace}
\newcommand{\mev}{\ensuremath{\mathrm{\,Me\kern -0.1em V}}\xspace}
\newcommand{\kev}{\ensuremath{\mathrm{\,ke\kern -0.1em V}}\xspace}
\newcommand{\ev}{\ensuremath{\mathrm{\,e\kern -0.1em V}}\xspace}
\newcommand{\gevc}{\ensuremath{{\mathrm{\,Ge\kern -0.1em V\!/}c}}\xspace}
\newcommand{\mevc}{\ensuremath{{\mathrm{\,Me\kern -0.1em V\!/}c}}\xspace}
\newcommand{\gevcc}{\ensuremath{{\mathrm{\,Ge\kern -0.1em V\!/}c^2}}\xspace}
\newcommand{\gevgevcccc}{\ensuremath{{\mathrm{\,Ge\kern -0.1em V^2\!/}c^4}}\xspace}
\newcommand{\mevcc}{\ensuremath{{\mathrm{\,Me\kern -0.1em V\!/}c^2}}\xspace}

\newcommand{\myconds}[1]{\langle#1\rangle}

\usepackage{cite} 
\usepackage{mciteplus}
\usepackage{tikz}
\usepackage{multirow}
\usepackage{ulem}

\begin{document}
\maketitle
\setcounter{tocdepth}{2}
\begin{abstract}

In \cite{Xiong:2022} we proposed a novel theoretical approach for non-perturbative QCD, named the inverse problem approach.
It extracts non-perturbative quantities by solving a dispersion integral equation as an inverse problem, using perturbative QCD and operator product expansion as inputs.
In this work, we apply this approach to the pion and $a_1(1260)$ decay constants, as well as the continuum spectral function.
The extracted $f_\pi = 130.8 \pm 8.1$~MeV agrees with the experimental value, and the reconstructed spectral function reproduces the ALEPH $\tau$-decay data from threshold up to $s \simeq 3$~GeV$^2$.
The advantages of this approach are as follows: 
(i) the whole non-perturbative range is solved simultaneously, yielding the ground state, excited states, and the continuum spectrum within a single framework; 
(ii) it is based on rigorous mathematics, with not only statistical but also systematic uncertainties well controlled and quantifiable; 
(iii) the regularized solutions exhibit convergence, allowing the precision to be systematically improved with better inputs; 
(iv) quark-hadron duality emerges naturally at high energy scales, avoiding the low-energy duality assumption and the Borel transformation, and eliminating their associated uncertainties.
Finally, the successful physical application of the inverse problem approach demonstrates its broad applicability to non-perturbative quantities.

\end{abstract}
\newpage
\tableofcontents
\section{Introduction}

How to calculate non-perturbative QCD quantities is a long-standing problem in particle physics and nuclear physics. 
In Ref.~\cite{Xiong:2022}, we constructed a novel and complete theoretical framework for such calculations, named the inverse problem approach. 
The idea was first proposed in Ref.~\cite{Li:2020xrz} to study $D^0-\overline D^0$ mixing. 
It determines the non-perturbative quantities in the dispersion integral equation as an inverse problem, using perturbative calculations as inputs. 
However, its realization is non-trivial. 
In Ref.~\cite{Xiong:2022}, we proved that the integral equation is ill-posed, i.e., the solution is unique but unstable. 
The uncertainty of the solution could be arbitrarily large even if the error of the perturbative inputs is small.  

Li proposed an inverse matrix method based on the expansion of Laguerre polynomials \cite{Li:2021gsx}, and has applied it to a wide range of physical problems, including hadron spectra, neutral meson mixing, and light-cone distribution amplitudes \cite{Li:2020fiz,Li:2020ejs,Li:2022qul,Li:2022jxc,Li:2024fko,Zeng:2026vyp,Mutuk:2025lak,Zhao:2024drr}.
The inverse matrix method is physically intuitive, yet its mathematical foundation remains unclear, which is particularly important for analyzing uncertainties.

Based on rigorous mathematics, we employ Tikhonov regularization to obtain stable approximate solutions in Ref.~\cite{Xiong:2022}. 
More importantly, as the error of the inputs $\delta\to 0$, the regularized solution converges to the true solution.
The framework in \cite{Xiong:2022} was tested only with toy models.
Here we apply it to a specific physical problem.
As the lightest hadron, the pion is easily produced in experiments.
Using the inverse problem approach, we calculate the pion decay constant, whose experimental measurement is very precise and is commonly used to test the applicability of non-perturbative QCD methods.

One of the advantages of the inverse problem approach is that it solves the entire non-perturbative range simultaneously, including the ground state, excited states, and the continuum spectrum.
By contrast, in many other methods one usually has to solve the ground state first and then the excited states one by one. 
As a result, the uncertainties of the excited states grow progressively larger, making reliable information on the excited states difficult to obtain. The situation is even worse for the continuum spectrum.
In this work, we simultaneously extract the decay constants of the pion and $a_1(1260)$, together with the continuum spectral function, from the axial-vector current.
Our result for the pion decay constant is in good agreement with the experiment.
The total spectrum also agrees well with the ALEPH data on $\tau$ decays in the axial-vector channel.
This agreement holds over the entire relevant range of the invariant mass squared, from $(3m_\pi)^2$ to $m_\tau^2\sim 3.16$~GeV$^2$, including both the $a_1(1260)$ peak and the continuum spectral function.
See Fig.~\ref{fig:final_result} and Table~\ref{tab:Results_Decay_constants}.

A key criterion for evaluating a theoretical method is whether it can estimate its systematic uncertainties, such as those arising from the assumptions or approximations it makes.  
Owing to its rigorous mathematical foundation, the inverse problem approach is capable of analyzing both the input-induced and method-intrinsic uncertainties, as shown in Table~\ref{tab:uncertainties}.  
Tikhonov regularization guarantees convergence of the regularized solution to the true solution as the input error $\delta \to 0$, with the regularization parameter $\alpha$ determined mathematically by the quasi-optimality criterion rather than by artificial choices.  
On the one hand, the approximation error arising from the finite value of $\alpha$ can be estimated via $\|R_\alpha Kf - f\|$, which represents a systematic uncertainty inherent to the method itself.  
On the other hand, the results are insensitive to all parameters appearing in the method, making it possible to estimate the systematic uncertainties arising from them.  
Furthermore, the convergence as the input error decreases ensures that the precision of the solution in the inverse problem approach can be systematically improved.

As a novel non-perturbative method, the inverse problem approach exhibits several distinctive features and yields interesting physical results.
It reproduces quark-hadron duality at high energy scales, while explicitly violating it at low-energy scales.
At scales larger than $\Lambda \sim 5\text{--}10~\mathrm{GeV}^2$, perturbative and OPE calculations are taken as input, whereas at scales smaller than $\Lambda$, non-perturbative quantities are to be solved.
As shown in Fig.~\ref{fig:final_result}, the solution from around $4~\mathrm{GeV}^2$ upward is very close to the OPE spectrum, indicating that duality is naturally restored at high energy scales.
The duality violation can also be estimated numerically to be about $(1\text{--}4)\%$ in the range $5\text{--}10~\mathrm{GeV}^2$.
Given the uncertainties in the OPE calculations, this is small, so quark–hadron duality is a good approximation at high energy scales.
In contrast, in the low-energy region, the duality is explicitly violated due to the existence of the pion pole and the $a_1(1260)$ resonance.
Unlike QCD sum rules where quark–hadron duality is assumed at low energy scales \cite{Shifman:1978bx,Shifman:1978by}, the inverse problem approach does not require the effective threshold parameter $s_0$, thus avoiding the associated uncertainties.
It also does not require Borel transformation, and our numerical results are indistinguishable with and without it.
Moreover, this approach can systematically analyze how different parts of the input affect the solution. For example, we can examine the contributions to $f_\pi$, $f_{a_1}$ and the continuum spectrum, from leading-order and next-to-leading-order perturbative calculations and from each condensate component, respectively.

The organization of this paper is as follows. 
The theoretical framework of the inverse-problem approach is introduced in Sec.~\ref{sec:Theoretical_framework}.
The numerical results are presented in Sec.~\ref{sec:numerical-results}. 
Physical implications are discussed in Sec.~\ref{sec:physical-discussions}.
A summary is given in Sec.~\ref{sec:summary}.

\section{Theoretical Framework}{\label{sec:Theoretical_framework}}

In this section, we present the theoretical framework for extracting the pion decay constant using the inverse problem approach.
This approach was originally introduced and systematically developed in Ref. \cite{Xiong:2022}.
It focuses primarily on the underlying mathematical formalism and employs three toy models to illustrate the effectiveness of regularization methods in addressing the ill-posedness of inverse problems.
In this work, we demonstrate the utility and potential of the inverse problem approach for non-perturbative QCD by applying it to calculate the decay constants of pion and excited states, as well as the continuum spectral function.

We begin with the correlation function used to extract the pion decay constant, along with its dispersion relation and the corresponding operator product expansion (OPE) input. 
Subsequently, we formulate the dispersion relation as an inverse problem, using the OPE input to reconstruct non-perturbative quantities including the pion decay constant and the spectral function.
Finally, we briefly outline the Tikhonov regularization procedure to solve the ill-posedness of the inverse problem.

\subsection{Correlation Function and Dispersion Relation}
The decay constant of pion $f_{\pi}$ is defined by the matrix element
\begin{equation}{\label{eq:definition_decay_constant}}
     \begin{aligned}
        \bra{0} j_{\mu 5}\ket{\pi } \equiv
        i f_{\pi} p_{\mu}^\pi \,,
     \end{aligned}
\end{equation}
where $j_{\mu 5}=\bar{u}\gamma_\mu\gamma_5 d$ denotes the axial-vector current and $p^{\pi}$ is the pion momentum.
As in QCD-based approaches, the determination of such quantities begins with the construction of the appropriate correlation functions. 
Based on Eq. \eqref{eq:definition_decay_constant}, we consider the following correlation function
\begin{equation}\label{eq:correlator}
    \Pi_{5\mu\nu}(q) = i \int d^4 x \, e^{i q \cdot x} \bra{0} T \left\{j_{\mu 5} (x) j_{\nu 5}^{\dagger}(0)  \right\} \ket{0} = -g_{\mu\nu} \Pi_g (q^2) + q^{\mu}q^{\nu} \Pi_q(q^2)\,,
\end{equation}
with the QCD non-perturbative vacuum $\ket{0}$. 

In the deep Euclidean region, $q^2 \ll 0$, the correlation function is calculable within the OPE.
By contrast, in the timelike region, $q^2>0$, the correlation function encodes hadronic information. 
Inserting a complete set $\sum_n\ket{n}\bra{n}$ of hadronic states that couple to $j_{\mu5}$, we obtain
\begin{equation}{\label{eq:correlator-im}}
    2 \operatorname{Im} \Pi_{5\mu\nu}(q) = \sum_n \bra{0}j_{\mu5} \ket{n} \bra{n} j_{\nu5}^{\dagger} \ket{0} d{\phi}_n (2\pi)^4\delta^{4}(q-p_n)\,,
\end{equation}
where $d\phi_n$ and $p_n$ denote the phase space and the momentum of state $\ket{n}$, respectively.
The details of the phase spaces and their integration can be seen in \cite{Colangelo:2000dp,Khodjamirian:2020btr}. 

For the axial-vector current $j_{\mu5}$, the complete set $\{\lvert n\rangle\}$ includes pseudoscalar states such as $\lvert\pi\rangle$, axial-vector states such as $\lvert a_1\rangle$, and continuum states such as $\lvert3\pi\rangle$ and $\lvert5\pi\rangle$.
Consequently, the correlator in Eq.~\eqref{eq:correlator-im} encodes both the pion decay constant $f_\pi$  defined in Eq.~\eqref{eq:definition_decay_constant}, and the decay constant $f_{a_1}$ of the axial-vector state, defined as follows
\begin{equation}{\label{eq:definition_decay_constant_a1}}
     \begin{aligned}  
        \bra{0} j_{\mu 5}\ket{a_1} \equiv \epsilon_{\mu}^{a_1} f_{a_1} M_{a_1}\,,
     \end{aligned}
\end{equation}
where $\epsilon_\mu^{a_1}$ and $M_{a_1}$ are the polarization vector and mass of $a_1$, respectively.

The pion matrix element in Eq.~\eqref{eq:definition_decay_constant} is proportional to the pion momentum $p_{\mu}$, while the axial-vector matrix element in Eq.~\eqref{eq:definition_decay_constant_a1} contains the polarization vector $\epsilon_{\mu}$. 
Consequently, the invariant amplitude $\Pi_g(q^2)$ in Eq.~\eqref{eq:correlator} with structure $g_{\mu\nu}$ includes only the contribution $a_1$, while $\Pi_q(q^2)$ receives contributions from both the pion and $a_1$.
To determine $f_\pi$ and compare it with experiment, we therefore focus our analysis on the invariant amplitude $\Pi_q(q^2)$.

\begin{figure} 
\centering 
\definecolor{darkpurple}{RGB}{72,40,120}
\usetikzlibrary{decorations.pathmorphing}
\begin{tikzpicture}[scale=1.1, transform shape=0.4]
    \fill[blue!20] (0,0) circle (2.5);
    \shade[shading=radial, inner color=green!90, outer color=blue!20] (0,0) circle (1.6);

    \draw[dashed, red] (0,0) circle (1.5);
     
    \draw[-, black, opacity=0.3] (-1.5,0) -- (0.3,0); 
    \draw[-, black, opacity=0.3] (0,-1.5) -- (0,1.5); 
    
     \draw[decorate, decoration={snake, amplitude=0.3mm, segment length=2.1mm},  line width=1.8pt] (1.50,0) -- (2.5,0); 
      \draw[->, opacity=0.3] (2.5,0) -- (2.8,0); 
       \node at (3.2,0) 
       {\small{$\textbf{Re}s$}};
        \node at (0,3.1) 
       {\small{$\textbf{Im}s$}};
     
     \draw[-, opacity=0.3] (-1.5,0) -- (-2.8,0); 
      \draw[decorate, decoration={snake, amplitude=0.3mm, segment length=2.2mm}, line width=1.8pt, blue] (0.3,0) -- (1.50,0); 
     \draw[->, opacity=0.3]  (0,1.5) -- (0, 2.8); 
     \draw[-,  opacity=0.3]  (0,-1.5) -- (0, -2.8); 
     
    \draw[->, blue, dashed] (0,0) -- (135:1.5);
    \node at (132:1) [above] {{\color{blue}\footnotesize{$\Lambda$}}};
     \node at (-0.0,1.9) {\small{\textrm{\textbf{pQCD and OPE}}}};
    \node at (0,-0.45) [below] {\small{\textrm{\textbf{Non-perturbative}}}}; 
    \node at (0,-0.78) [below] {\small{\textrm{\textbf{QCD}}}};
    
    \draw[thick] (-1.8,0) circle (1.5pt); 
    \node at (-1.75,0.3) {{\color{black}\footnotesize{$q^2$}}};
    \draw[thick] (-2.2,0) circle (1.5pt); 
    \node at (-2.15,0.25) {{\color{black}\footnotesize{$\sigma_0$}}};
    
    \node at (0.9,0.3) {{\color{blue}\footnotesize{$\textbf{Im} \Pi(s)$}}};
    

\filldraw[black,blue] (0.3, 0) circle (0.8pt);            
\node at  (0.3, -0.22) {\small{$t_{\min}$}}; 

        \draw[line width=1.5pt,darkpurple!80] (0.30, 0.14) arc (-90:90:-0.14);
         \draw[-,line width=1.5pt,darkpurple!80] (0.30,0.14)--(2.49,0.14); 
         \draw[-,line width=1.5pt,darkpurple!80] (2.49,-0.14)--(0.30,-0.14);
         \draw[->,line width=1.5pt,darkpurple!80] (2.49, -0.14) -- (1.5,-0.14); 
         \draw[->,line width=1.5pt,darkpurple!80] (0.30, 0.14) -- (1.65,0.14);
     \draw[->, line width=1.5pt,darkpurple!80]  (1.80,1.74) -- (1.7678, 1.7678);
        \draw[line width=1.5pt,darkpurple!80] (2.495,0.12) arc (2.5:357:2.5);
 \draw[->, line width=1.5pt,darkpurple!80]  (-1.80,-1.74) -- (-1.7678, -1.7678);
      \node at (1.0,-1.9) {\small{\textit{\textbf{\textcolor{darkpurple!80}{$\left| s\right|=R$}}}}};
\end{tikzpicture} 
\caption{Complex $s$-plane for the subtracted dispersion relation and its inverse problem. The integration contour consists of two horizontal lines running above and below the physical branch cut, connected by a large circle at infinity. The scale $\Lambda$ marks the boundary between the perturbative and non-perturbative QCD regimes. The imaginary part of correlation function, $\operatorname{Im} \Pi(s)$, is the target of the inverse problem, and its integration interval is taken from $t_{\rm min}$ to $\Lambda$. The input point $q^2$ and the subtraction point $\sigma_0$ are both placed in the deep Euclidean region to ensure the validity of the OPE.} \label{fig:Contour} 
\end{figure}
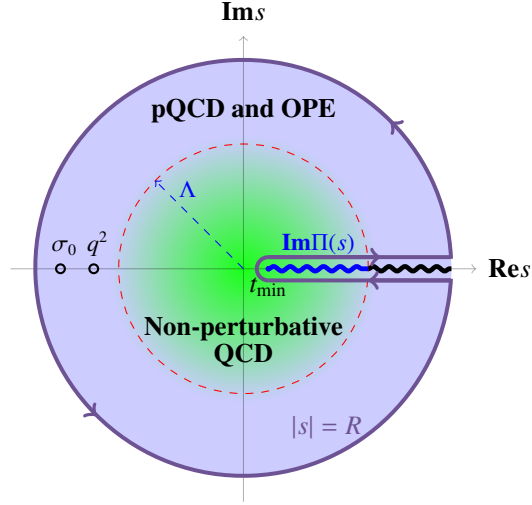

Applying Cauchy's theorem to the contour in Fig.~\ref{fig:Contour} and introduced in Ref.~\cite{Xiong:2022}, we obtain the dispersion relation for $\Pi_q(q^2)$,
\begin{equation}{\label{eq:dispersion_relation1}}
    \Pi_q(q^2) = \frac{1}{\pi} \int_{t_{\rm min}}^{\infty} \,ds\frac{\operatorname{Im} \Pi_q(s) }{s-q^2} + \frac{1}{2 \pi i} \oint_{|s|=R \to \infty}\,ds \frac{\Pi_q(s)}{s-q^2}.
\end{equation}
It relates the OPE representation of $\Pi_q(q^2)$ in the deep Euclidean region ($q^2\ll 0$) to its spectral function in the physical timelike region ($q^2>0$).
If $\Pi_q(s)$ vanishes as $|s|\to\infty$, the second term on the right-hand side of Eq.~\eqref{eq:dispersion_relation1} vanishes, yielding the standard unsubtracted dispersion relation as shown in Ref.~\cite{Xiong:2022}.

For large $|s|$, the asymptotic behavior of $\Pi_q(s)$ is governed by its OPE and perturbative calculations.
The OPE is truncated up to the dimension-six operator, and the perturbative corrections are computed up to the next-to-leading-order (NLO) in $\alpha_s$ ~\cite{Colangelo:2000dp}, 
%
%
\begin{equation}{\label{eq:OPE_calculation}}
    \begin{aligned}
     \Pi_{q}(q^2)  =& - \frac{1}{4\pi^2} \left(1+ \frac{\alpha_s(\mu)}{\pi} \right)
         \ln\left( \frac{-q^2 }{\mu^2} \right)  +\frac{(m_u  + m_d) \myconds{\bar{q}q}}{(q^2)^2}  + \frac{1}{12\pi} \frac{\myconds{\alpha_s G^2}}{(q^2)^2} -\frac{352 \pi}{81}  \frac{\alpha_s(\mu) \myconds{\bar{q}q}^2}{(q^2)^3}\,.
    \end{aligned}
\end{equation}
The leading-logarithmic behavior indicates that $\Pi_q(s)$ grows as $\ln(-s)$ for large  $|s|$, rather than vanishing in this limit.
Hence, the unsubtracted dispersion relation in Eq.~\eqref{eq:dispersion_relation1} is divergent, and a subtraction is necessary to render the large-circle contribution vanishing.

For a subtraction at a fixed point $\sigma_0$ with $\sigma_0<t_{\rm min}$, the Cauchy integral formula gives
\begin{equation}{\label{eq:dispersion_relation_sigma0}}
    \Pi_q(\sigma_0) = \frac{1}{\pi} \int_{t_{\rm min}}^{\infty} \,ds\frac{\operatorname{Im} \Pi_q(s) }{s-\sigma_0} + \frac{1}{2 \pi i} \oint_{|s|=R \to \infty}\,ds \frac{\Pi_q(s)}{s-\sigma_0}\,.
\end{equation}
Subtracting Eq.~\eqref{eq:dispersion_relation_sigma0} from Eq.~\eqref{eq:dispersion_relation1} yields
\begin{equation}
    \frac{\Pi_q(q^2)-\Pi_q(\sigma_0)}{q^2 - \sigma_0} = \frac{1}{\pi}\int_{t_{\rm min}}^{\infty}\,ds \frac{\operatorname{Im} \Pi_q(s)}{(s-q^2)(s-\sigma_0)} + \frac{1}{2 \pi i} \oint_{|s|=R \to \infty}ds \frac{\Pi_q(s)}{(s-q^2)(s-\sigma_0)}\,,
\end{equation}
where $\sigma_0$ is chosen in the deep Euclidean region, satisfying $\sigma_0<q^2\ll0$, so that $\Pi_q(\sigma_0)$ can be calculated perturbatively.
Due to the extra factor $(s-\sigma_0)^{-1}$ in the integrand, the large-circle contribution vanishes even though $\Pi_q(s)$ itself grows logarithmically.
We then obtain the subtracted dispersion relation
\begin{equation}{\label{eq:dispersion_relation_subtracted}}
\begin{aligned}
    \frac{\Pi_{q}(q^2) - \Pi_{q}(\sigma_0)}{q^2-\sigma_0}&=  \frac{1}{\pi}\int_{t_{\rm min}}^{\infty} ds \frac{\operatorname{Im}\Pi_{q}(s)}{(s-\sigma_0)(s-q^2)}\,.
\end{aligned}
\end{equation}

For $s\gg0$, it is reliable to approximate $\operatorname{Im}\Pi_{q}(s)$ by  the OPE result $\operatorname{Im}\Pi_{q}^{\rm OPE}(s)$.
The imaginary part arises from the analytic continuation of the logarithm,
$\ln\left(-s/\mu^2\right)\to \ln\left(s/\mu^2\right) -\pi i$. 
Eq.~(\ref{eq:OPE_calculation}) then yields 
\begin{equation}\label{eq:perturbative_imaginary_Pi_q}
    \operatorname{Im}\Pi_{q}^{\rm OPE}(s)= \frac{1}{4 \pi}\left(1+ \frac{\alpha_s(\mu)}{\pi}\right)\,, \quad {\rm for}\quad s\gg 0\,.
\end{equation}
This will serve both as the boundary condition at $\Lambda$ in the inverse problem approach and as a benchmark for comparison with the solution.

\subsection{Inverse Problem of Dispersion Relation}

Following the inverse problem approach in Ref.~\cite{Xiong:2022} and Fig.~\ref{fig:Contour}, we separate the dispersion integral in Eq.~(\ref{eq:dispersion_relation_subtracted}) at a scale $\Lambda$, with the regions below and above $\Lambda$ identified as non-perturbative and perturbative, respectively. 
Eq.~\eqref{eq:dispersion_relation_subtracted} can then be rewritten as an integral equation of the inverse problem,
\begin{equation}{\label{eq:original_inverse_problem}}
\begin{aligned}
   \int_{t_{\rm min}}^{\Lambda} ds \frac{\rho(s)}{(s-\sigma_0)(s-q^2)} 
    & =  \frac{\Pi_{q}(q^2) - \Pi_{q}(\sigma_0)}{q^2-\sigma_0} - \int_{\Lambda}^{\infty} ds \frac{\rho^{\rm OPE}(s)}{(s-\sigma_0)(s-q^2)} \,,  
\end{aligned}
\end{equation}
where 
\begin{equation}
\rho(s)\equiv \frac{1}{\pi}\operatorname{Im}\Pi_q(s) ~~~\text{and}~~~ \rho^{\rm OPE}(s) \equiv \frac{1}{\pi}\operatorname{Im}\Pi_q^{\rm OPE}(s)
\end{equation}
denote the full and OPE spectral functions, respectively.
The spectral function $\rho(s)$ over $t_{\rm min}\leqslant s\leqslant\Lambda$, appearing on the left-hand side, is the non-perturbative quantity to be reconstructed.
By contrast, all quantities on the right-hand side are theoretically calculable: $\Pi_q(q^2)$ and $\Pi_q(\sigma_0)$ are evaluated from the OPE in the deep Euclidean region, $q^2,\sigma_0\ll0$, while quark-hadron duality is reliable for $s>\Lambda$.
Therefore, the central task is to identify the low-energy non-perturbative spectral function from the known OPE input by solving this integral equation.
This is a Fredholm integral equation of the first kind, which determines the unknown integrand from the known integral result, and thus constitutes a typical inverse problem.

Before proceeding, we discuss the admissible ranges of the parameters $\Lambda$ and $\sigma_0$.
The separation scale $\Lambda$ is not uniquely defined, since the transition between the perturbative and non-perturbative regimes is continuous rather than abrupt.
Therefore, physical results should  be stable against variations of $\Lambda$ within a reasonable range.
The numerical analysis in the next section confirms this expectation.
Nevertheless, $\Lambda$ cannot be arbitrarily large or small.
If $\Lambda$ is too large, the reconstruction interval $[t_{\rm min},\,\Lambda]$ becomes wider, increasing the amount of information to be determined, and thus reducing the quality of reconstruction.
Conversely, if $\Lambda$ is too small, the perturbative expansion and the OPE become less convergent, increasing the importance of higher-order and higher-power corrections, ultimately degrading the reconstruction quality.

We now turn to the determination of the subtraction point $\sigma_0$.
The subtraction ensures the convergence of the dispersion integral.
However, the subtraction introduces $\Pi_q(\sigma_0)$ as an additional input.
Its reliable evaluation requires $\sigma_0$ to lie sufficiently deep in the Euclidean region as seen in Fig.~\ref{fig:Contour}, where the OPE exhibits good convergence.
Moreover, physical observables should be independent of the arbitrary choice of subtraction point. 
As will be shown in the numerical results, the extracted results exhibit negligible dependence on $\sigma_0$.
In QCD sum rules, the subtraction point is typically chosen as $\sigma_0=0$. 
The subtraction constant $\Pi(0)$ vanishes identically in the vector channel, whereas in the axial-vector channel it receives non-perturbative contributions and is generally nonzero. 
Treating $\Pi_q(0)$ as an additional unknown would weaken the constraints on the inverse problem and thus reduce the reliability of extraction.
The deep-Euclidean subtraction point adopted here avoids this complication.

Eq.~\eqref{eq:original_inverse_problem} provides the general form of the inverse problem integral equation derived from a subtracted dispersion relation.
For the determination of $f_\pi$, the pion contribution to the spectral function takes the form of a delta-function pole at $s=m_\pi^2$.
However, directly reconstructing this delta-function contribution is a highly challenging task.
The objective here is to determine the properties of the known pion state, such as its decay constant, light-cone distribution amplitude (LCDA), and parton distribution function (PDF), rather than to establish the existence of the state itself. 
Since its mass $m_\pi$ and delta-function pole structure are already known experimentally, these can be incorporated as prior information in the inverse problem.
Incorporating this prior information, we decompose the full spectral function into an isolated pion-pole contribution and a continuum contribution,
\begin{equation}{\label{eq:spectral_decompose_N1}}
    \rho(s) = f_\pi^2 \delta(s-m_\pi^2) + \rho^c(s) \theta(s-s_{\rm th})\,.
\end{equation}
The continuum spectral function $\rho^c(s)$ accounts for all higher-mass excited states and multi-particle contributions, and is assumed to be a smooth function above the threshold in the numerical analysis.
$s_{\rm th}$ denotes the physical hadronic threshold of the continuum spectral density, e.g. $s_{\rm th}=(3m_\pi)^2$ for the axial-vector current channel.

The decomposition of the spectral function into a pole and a continuum, as in Eq.~\eqref{eq:spectral_decompose_N1}, is a standard ansatz in QCD sum rules~\cite{Shifman:1978bx,Shifman:1978by}. 
However, the key difference in the inverse problem approach is that the continuum spectral function $\rho^c(s)$ is treated as unknown to be solved, whereas in conventional QCD sum rules it is eliminated via the assumption of global quark-hadron duality. 
This distinction, along with its implications, will be discussed in detail in Sec.~\ref{sec:quark-hadron-duality}. 
The treatment of $\rho^c(s)$ as a dynamical quantity naturally requires boundary conditions: at the threshold $s_{\rm th}$, the continuum density must vanish $\rho^c(s_{\rm th})=0$, while at the upper endpoint $\Lambda$, the quark-hadron duality ansatz imposes that the density is equal to the OPE one. 
Note that the full spectral function starts at $t_{\rm min}=m_\pi^2$, while the continuum threshold is $s_{\rm th}=(3m_\pi)^2$ for the axial-vector channel.

We emphasize that the explicit delta function in Eq.~\eqref{eq:spectral_decompose_N1} serves primarily to make the physical interpretation transparent by isolating the pion-pole contribution. 
However, this decomposition is not a result extracted from the numerical inversion of Eq.~\eqref{eq:original_inverse_problem}. 
Rather, it is prior information that we impose based on the known existence and mass of the pion. 

In Ref.~\cite{Xiong:2022}, we have proved mathematically that the inverse problem associated with the dispersion relation is ill-posed in the $L^2$ space.
However, the delta distribution appearing in Eq.~\eqref{eq:spectral_decompose_N1} does not belong to $L^2$, since it is not square integrable.
For the actual mathematical implementation, we substitute Eq.~\eqref{eq:spectral_decompose_N1} into Eq.~\eqref{eq:original_inverse_problem} and carry out the integration over the delta function. 
This yields the working equation of our inverse problem,
\begin{equation}\label{eq:final_dispersion_relation}
\begin{aligned}
     \frac{f_{\pi}^2}{(m_\pi^2-\sigma_0)(m_\pi^2-q^2)}+ \int_{s_{\rm th}}^\Lambda ds \frac{\rho^c(s)}{(s-\sigma_0)(s-q^2)} =\frac{\Pi_q(q^2) - \Pi_{q}(\sigma_0)}{(q^2 -\sigma_0)} - \int_{\Lambda}^{\infty} ds \frac{\rho^{\rm OPE}(s)}{(s-\sigma_0)(s-q^2)} \,.
\end{aligned}
\end{equation}
At this point, the mathematical structure of the problem has changed compared to that defined by Eq.~\eqref{eq:original_inverse_problem}. 
In Eq.~\eqref{eq:original_inverse_problem}, the sole unknown is the continuous function $\rho(s)$ throughout the interval $[t_{\rm min},\Lambda]$, which is a standard Fredholm problem of the first kind. 
In contrast, Eq.~\eqref{eq:final_dispersion_relation} involves two distinct unknown components: the continuum spectral function $\rho^c(s)$ in $s_{\rm th}\leqslant s\leqslant \Lambda$ and the isolated parameter $f_\pi^2$. 
Consequently, the inversion problem is no longer a pure function-reconstruction task, but becomes a mixed determination problem that simultaneously solves for a function and a parameter. 
This reformulation changes the mathematical problem, necessitating a new analysis of the properties of the integral equation.

To facilitate the analysis of its mathematical structure, we rewrite Eq.~\eqref{eq:final_dispersion_relation} in a more compact form.
Defining
\begin{equation}\label{eq:N-dispersion_relation_dimensionlessnext}
       g(q^2) \equiv \frac{\Pi_q(q^2) - \Pi_{q}(\sigma_0)}{q^2 -\sigma_0} - \int_{\Lambda}^{\infty}  \frac{\rho^{\rm OPE}(s)\, ds}{(s-\sigma_0)(s-q^2)} \,,
\end{equation}
Eq.~\eqref{eq:final_dispersion_relation} can be simply expressed as
\begin{equation}{\label{eq:final_equation}}
\begin{aligned}
    \frac{f_{\pi}^2}{(m_\pi^2-\sigma_0)(m_\pi^2-q^2)}+ \int_{s_{\rm th}}^\Lambda ds \frac{\rho^c(s)}{(s-\sigma_0)(s-q^2)} = g(q^2)\,,
\end{aligned}
\end{equation}
where $q^2\in[q_{\rm low}^2,q_{\rm up}^2]$, $\sigma_0< q_{\rm low}^2<q_{\rm up}^2\ll0$, $s\in[s_{\rm th},\Lambda]$, $0<m_\pi^2<s_{\rm th}<\Lambda$. 
The inverse problem aims to determine $f_\pi$ and $\rho^{c}(s)$ with the given input $g(q^2)$. 

The main challenge in solving the above problem lies in its ill-posed nature. 
As rigorously established in \cite{Xiong:2022}, the inverse problem for the dispersion relation is ill-posed.
Although its solution is unique, it is unstable against small perturbations in the input data, which can produce arbitrarily large errors in the reconstructed solution. 
The inverse problem defined by Eq.~\eqref{eq:final_equation} differs from the original formulation in \cite{Xiong:2022} in two respects: the inclusion of an isolated term $f_\pi^2$, and an integral kernel with a quadratic denominator, $1/[(s-q^2)(s-\sigma_0)]$. 
In this work, we show that this modified problem is also ill-posed, admitting a unique but unstable solution. 
Consequently, regularization is essential to ensure numerical stability.
The regularization method is introduced below, and its numerical implementation is presented in the next section.
For the convenience of readers focused on the physical application of the inverse problem, a detailed proof of the ill-posedness is provided in Appendix~\ref{sec:proof_illposedness}.

\subsection{The Inverse Problem Approach}
The essential idea of regularization is to replace the original ill-posed operator by a family of well-posed approximations. 
Following Ref.~\cite{Xiong:2022}, we adopt the Tikhonov regularization, which provides particularly transparent mathematical properties. 
The regularized solution is guaranteed to converge to the true solution as the noise level vanishes. 
It also provides a rigorous basis for quantifying the input-induced and method-intrinsic uncertainties, and enables the reconstruction precision to be improved in a controlled manner. 
These advantages will be demonstrated by the numerical results in the next section.

We now briefly outline the numerical procedure for the inverse problem based on Eq.~\eqref{eq:final_equation}.
The Tikhonov functional in this problem is defined as 
\begin{equation}\label{eq:Tikhonov_functional}
J_\alpha(f)=\left\|Kf -g^{\delta} \right\|^2_G+ \alpha \left\|\rho^c \right\|^2_F,
\end{equation}
where $F=L^2(s_{\rm th},\Lambda)$ and $G=L^2(q_{\rm low}^2,q_{\rm up}^2)$.
Note that the solution $f$ incorporates the pole contribution.
Since $g(q^2)$ has dimension $-2$, it follows that $\rho^c(s)$ is dimensionless. 
We therefore construct a dimensionless solution vector, $f=(f_\pi^2/m_\pi^2, \rho^c)$, with 
$Kf$ denoting the left-hand side of Eq.~(\ref{eq:final_equation}).
The regularized approximate solution is then given by the minimizer of this functional,
\begin{equation}\label{eq:f_alpha^delta}
f_\alpha^\delta =\operatorname*{arg\,min}_{f}J_\alpha(f).
\end{equation}

The first term in Eq.~(\ref{eq:Tikhonov_functional}) is the residual, which quantifies the agreement between the reconstructed solution and the input data, analogously to the conventional $\chi^2$ fit.
The second term is the penalty, where the regularization parameter $\alpha$ stabilizes the solution of the ill-posed problem against overfitting.
For the present problem, the instability arises primarily from reconstruction of the continuum spectral contribution.
We therefore apply regularization exclusively to $\rho^c$ via the penalty term $\alpha\lVert \rho^c\rVert^2_F$, which suppresses spurious high-frequency oscillations and enforces smoothness in $\rho^c$.

The solution space for $\rho^c$ is taken to be the standard $L^2$ space in the proof of ill-posedness presented in Appendix~\ref{sec:proof_illposedness}, for mathematical generality. 
However, as discussed in \cite{Xiong:2022}, when the smoothness of the solution is desired in practice, the $H^1$ space is a more suitable choice for $\rho^c$. 
Recall that $L^2$ consists of square integrable functions, while $H^1$ additionally requires square integrable for their first weak derivatives. 
It is worth noting that the ill-posedness established in the $L^2$ setting also persists in the $H^1$ space, as the latter is continuously embedded in the former.
Hence the regularization strategy adopted in the $H^1$ framework remains well motivated.
For completeness, we refer the reader to the Appendix of Ref.~\cite{Xiong:2022} for detailed definitions of these function spaces.

To minimize the functional $J_\alpha$, two steps are involved. 
First, the function $\rho^c(s)$ is discretized using the finite-element method $\rho^c(s)=\sum_ic_i\varphi_i(s)$, where $\varphi_i(s)$ are the hat functions. 
Second, the minimization is achieved by setting the partial derivative of $J_\alpha$ with respect to each coefficient $c_i$ to zero. 
Due to the linearity of all problems considered here, this procedure reduces to solving a linear system.
The implementation details are provided in Section~4.4 of Ref.~\cite{Xiong:2022}.
However, the additional pion-pole contribution introduces a modification to the original formalism. 
To clarify this change, we provide the corresponding explicit expression in Appendix~\ref{sec:Minimization_functional}.

Finally, we obtain the regularized solution $f_\alpha^\delta$, which depends on the regularization parameter $\alpha$ and the input error $\delta$.
As a mathematical parameter, $\alpha$ should not significantly affect the underlying physical results. 
Hence, the solution is expected to exhibit a stability plateau with respect to $\alpha$.
To determine $\alpha$, the quasi-optimality criterion is adopted, 
\begin{equation}
\alpha_{\rm opt}= \underset{\alpha}{\arg \min} \left\| \alpha\frac{\partial f_{\alpha}^c(s)}{\partial \alpha} \right\|.
\end{equation}
Minimizing the partial derivative of $f_{\alpha}^c(s)$ with respect to $\alpha$ effectively identifies a plateau where $f_\alpha^\delta$ is stable against variations in $\alpha$.
This plateau will be demonstrated numerically below.

As established in Ref.~\cite{Xiong:2022}, under suitable a priori conditions, the Tikhonov regularized solution $f_{\alpha}^\delta$ converges to the exact solution $f$ as $\delta\to0$.
This convergence is essential because it ensures that the inverse problem approach can be systematically improved by reducing the input error $\delta$.
Moreover, this approach provides a numerical framework for quantifying the input-induced and method-intrinsic uncertainties, which also diminish as $\delta$ decreases.
These uncertainties will be further examined in the numerical analysis in the next section.
A more comprehensive discussion of the underlying principles and regularization methods can be found in Ref.~\cite{Xiong:2022}.

We now summarize the criteria for assessing the quality of a reconstructed solution.
A reliable solution must simultaneously satisfy two conditions: stability plateaus with respect to both $\alpha$ and $\Lambda$.
The $\alpha$-plateau criterion reflects that the physical results should be insensitive to variations of this mathematical parameter.
The $\Lambda$-plateau criterion, on the other hand, requires that the reconstructed spectral function be consistent with its perturbative QCD counterpart near the upper boundary $s\simeq\Lambda$.

\section{Numerical Results}\label{sec:numerical-results}

In this section, we present the numerical results obtained via the inverse problem approach with Tikhonov regularization. 
The solutions for decay constants and spectral density are stable and agree well with the experimental data. 
In addition, a thorough analysis of uncertainties is provided. 
Finally, it is demonstrated that increasing the precision of the input data leads to convergent solutions, thereby confirming that the accuracy of the inverse-problem solution can be systematically improved.

\subsection{Results under Tikhonov Regularization}

The OPE calculation for the correlation function in the deep Euclidean region, given in Eq.~\eqref{eq:OPE_calculation}, involves various vacuum condensates. For our numerical analysis, we adopt the values listed below, which are taken from Refs.~\cite{Narison:2014wqa,FlavourLatticeAveragingGroup:2019iem},
%
%
\begin{equation}{\label{eq:input_condensates}}
    \begin{aligned}
\begin{array}{r@{\;}l@{\qquad}r@{\;}l}
  m_u &= 2.16 \pm 0.04\, \left[\frac{\alpha_s(\mu)}{\alpha_s(2\,\text{GeV})} \right]^{4/\beta_0} [\mathrm{MeV}]\,, &
  &\myconds{\bar q q} = -(0.272 \pm 0.005)^3 \left[\frac{\alpha_s(\mu)}{\alpha_s(2\,\text{GeV})} \right]^{-4/\beta_0}\,[\mathrm{GeV}]^3\,,\\
  m_d &= 4.70 \pm 0.04\, \left[\frac{\alpha_s(\mu)}{\alpha_s(2\,\text{GeV})} \right]^{4/\beta_0} [\mathrm{MeV}]\,,
  &
  &\myconds{\alpha_s G^2} = 0.07 \pm 0.02\,[\mathrm{GeV}]^4\,.
\end{array}
    \end{aligned}
\end{equation}
where the scale dependence of the relevant condensates is explicitly included, with $\beta_0 = 11- \frac{2}{3}n_f$ and $n_f$ the number of active quark flavors at scale $\mu$.
We note that most of these condensate values are extracted from QCD sum rules and thus come with additional uncertainties that cannot be reliably quantified and hence are not accounted for here.
In practice, their true uncertainties may be larger than those quoted in Eq.~\eqref{eq:input_condensates}.
In this work, we take these values as given.
Future determinations using the inverse problem approach, as more observables become available, may provide improved extractions of these condensates.
In addition, the parameters in the integral equation Eq.~(\ref{eq:final_equation}) are set as $\Lambda=6$ GeV$^2$, $\sigma_0=-400$ GeV$^2$, $q_{\rm low}^2=-300$ GeV$^2$, $q_{\rm up}^2=-6$ GeV$^2$, and the discretization interval of $\rho^c$ is taken as $h=0.01$ GeV$^2$.

\begin{figure}
                \centering
                 \subfigure[\rm Without regularization]{\includegraphics[width=0.315\textwidth ]{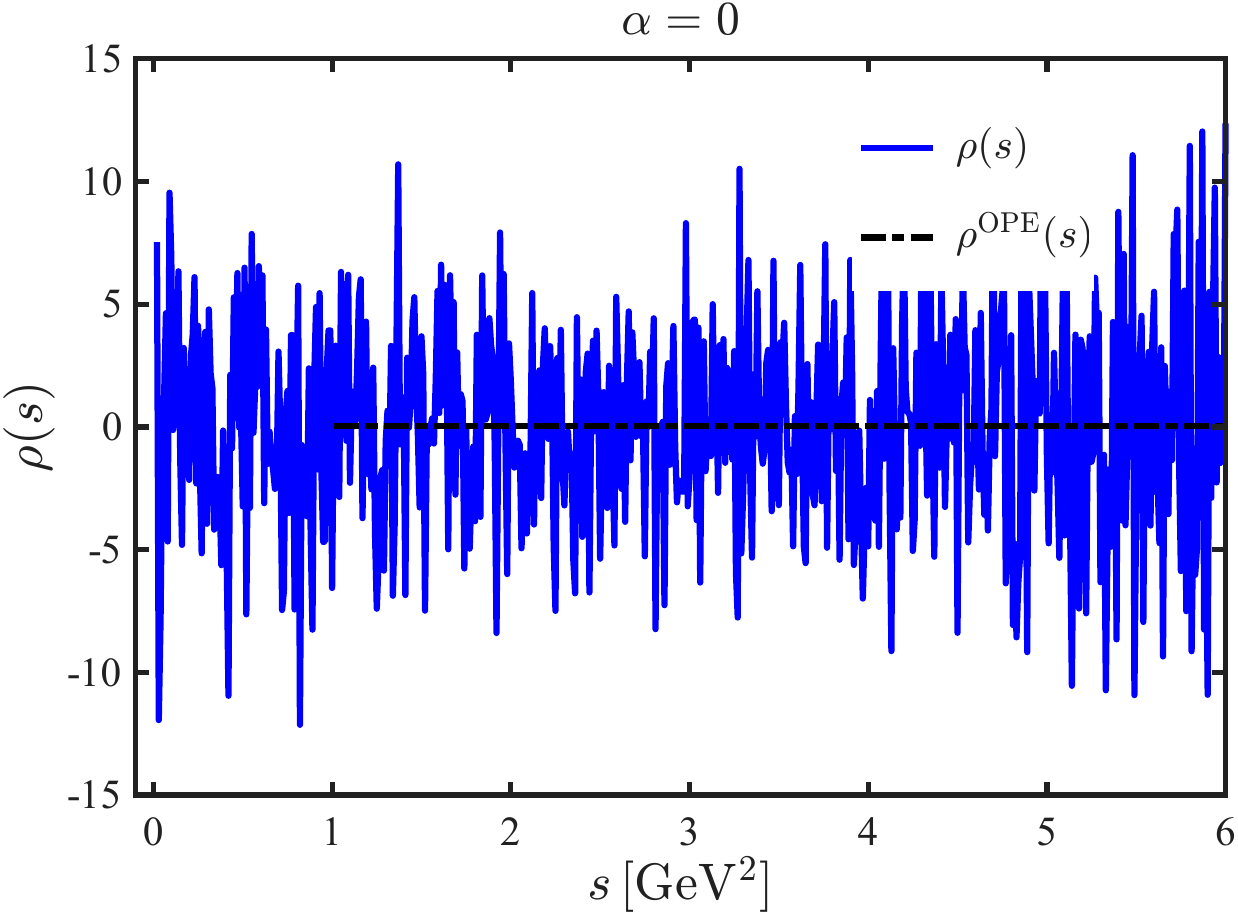}}
                 \hspace{2mm}
                 \subfigure[\rm Under regularization]{\includegraphics[width=0.315\textwidth ]{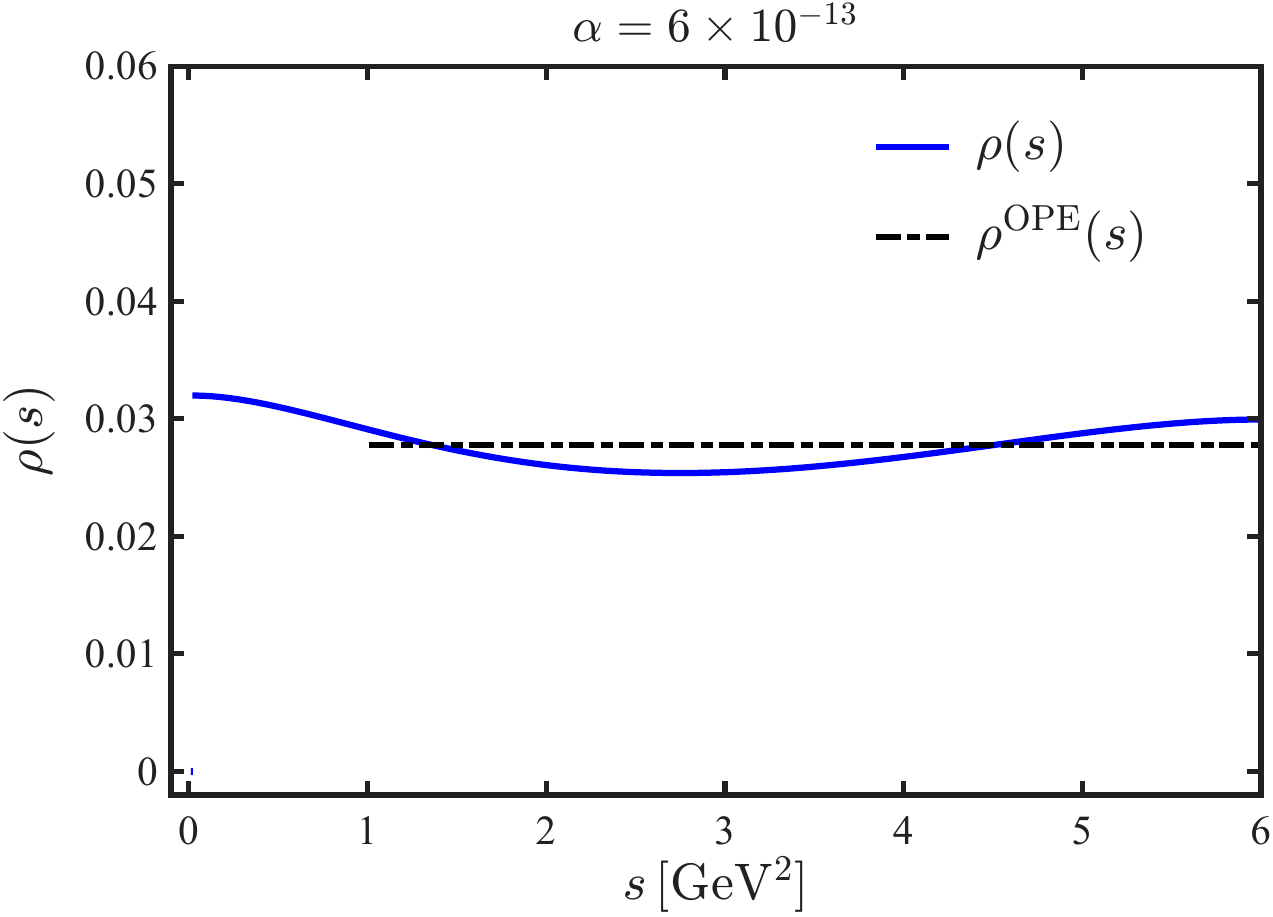}}
                 \hspace{2mm}
                \subfigure[\rm + boundary conditions]{\includegraphics[width=0.315\textwidth ]{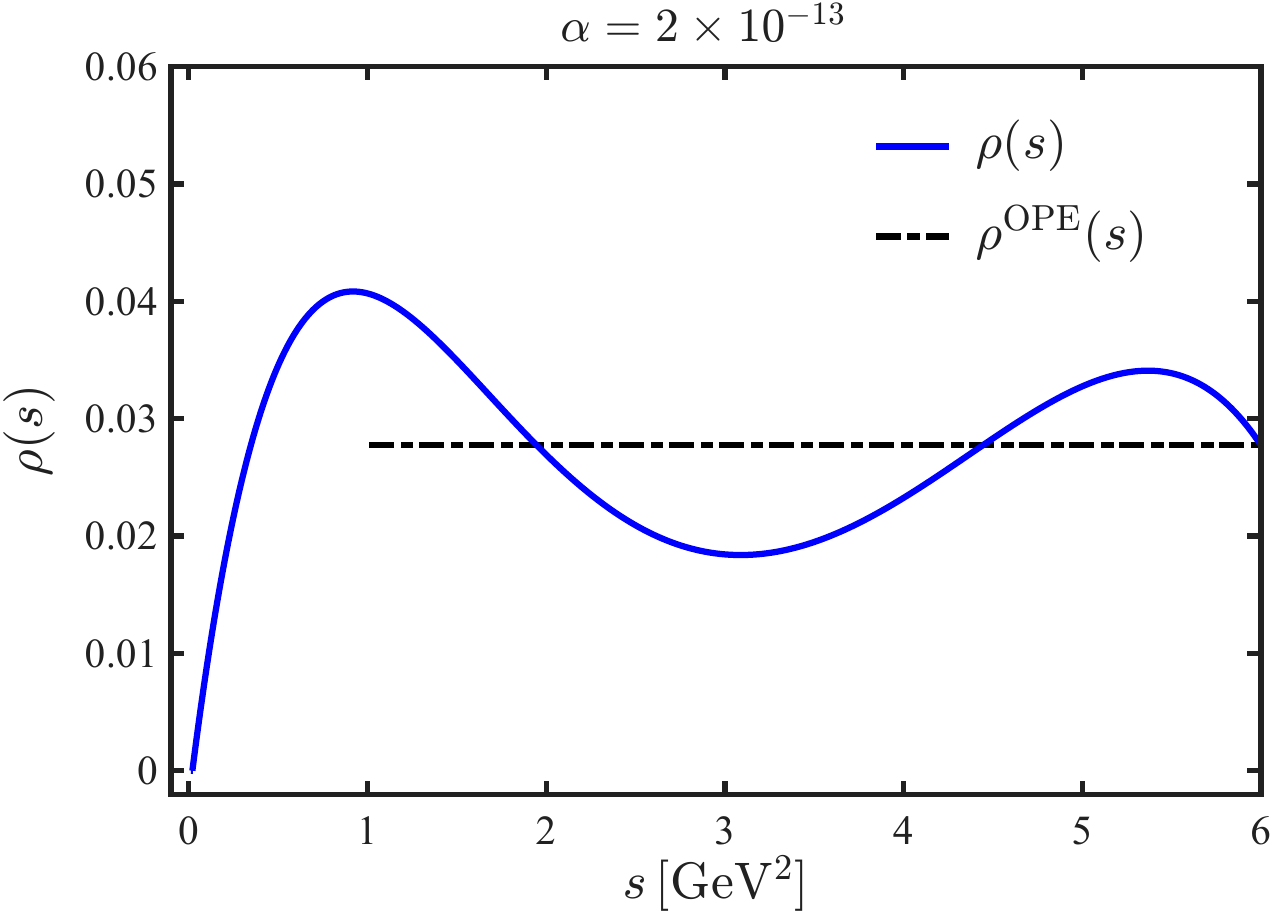}}

               \subfigure[\rm + pion pole]{\includegraphics[width=0.315\textwidth ]{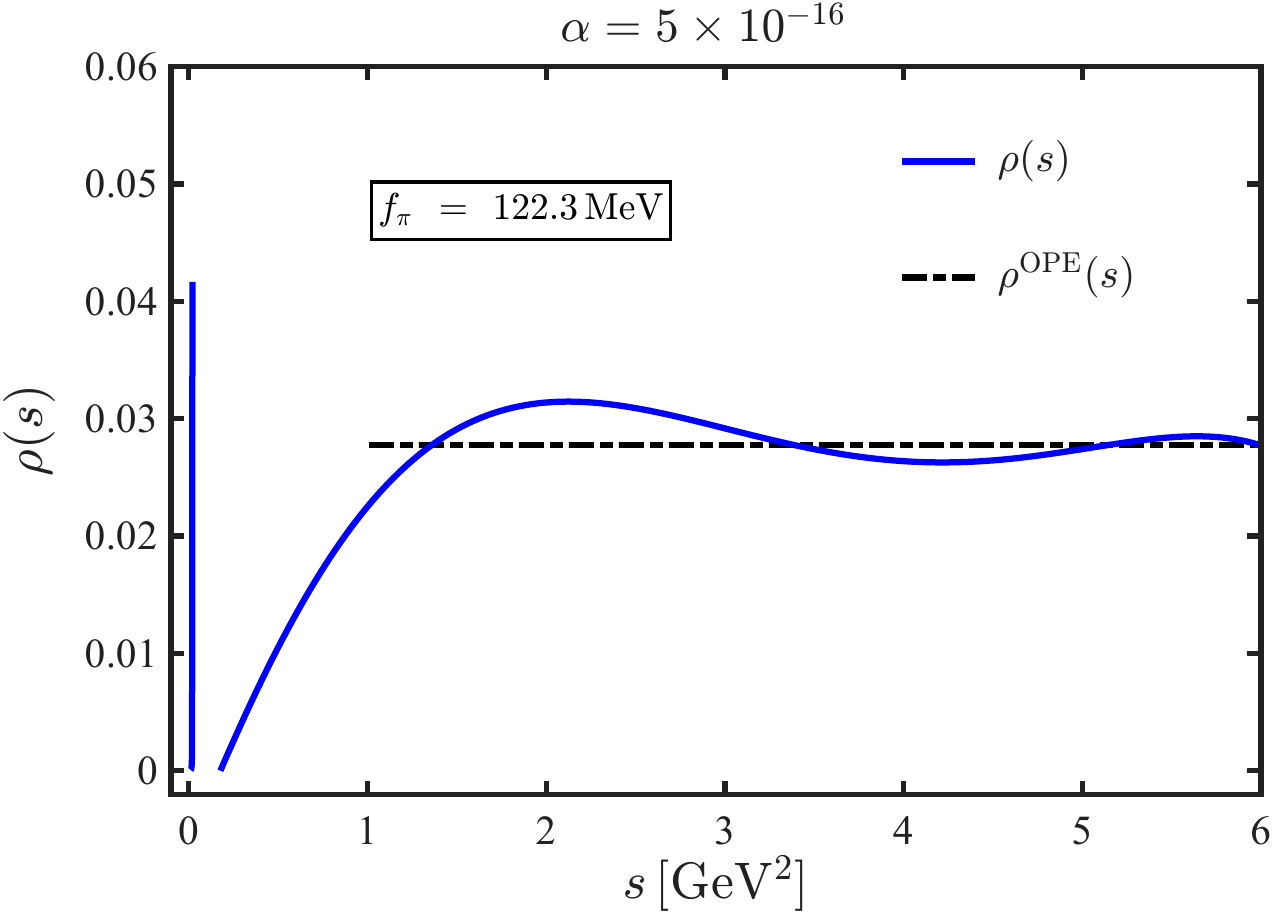}}
               \hspace{2mm}
               \subfigure[\rm + $a_1(1260)$ resonance]{\includegraphics[width=0.315\textwidth ]{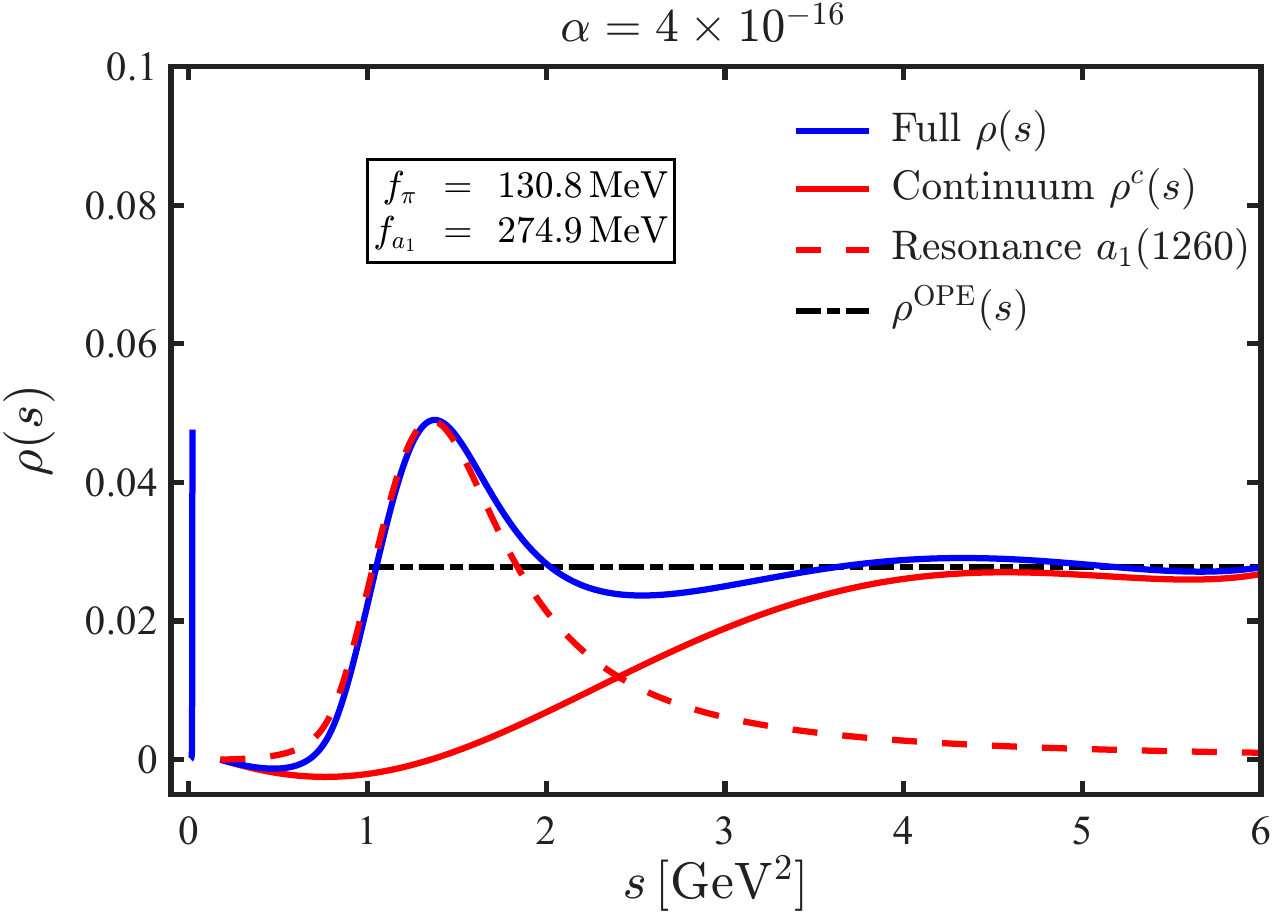}}
               \hspace{2mm}
               \subfigure[\rm + $a_1(1640)$ resonance]{\includegraphics[width=0.315\textwidth ]{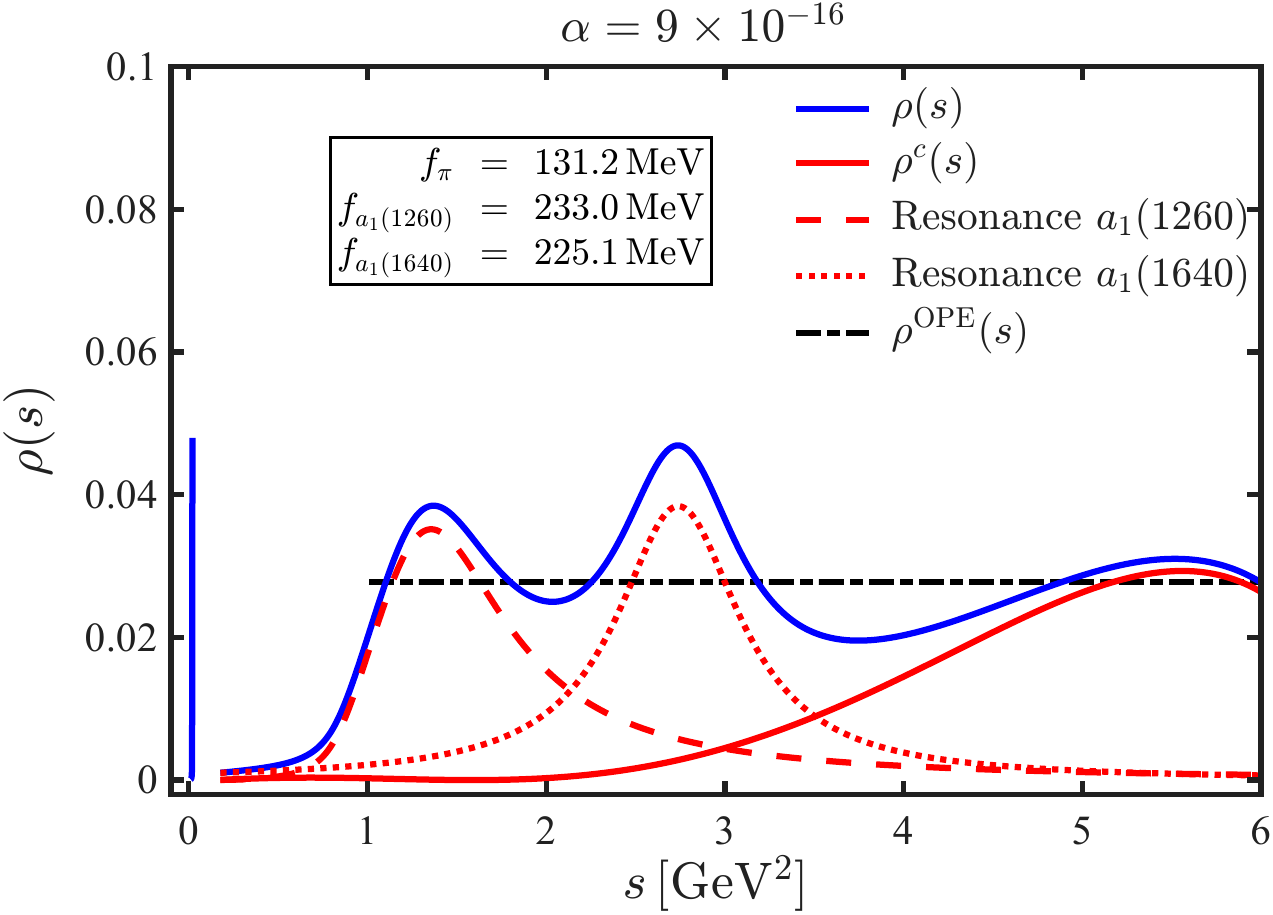}}
            \caption{Numerical results. (a) Result without regularization. (b)–(f) Results under Tikhonov regularization with progressive incorporation of priors: (b) baseline without any prior; (c) with boundary conditions; (d) incorporating the pion pole; (e) augmented with the $a_1(1260)$ resonance; (f) further augmented with the $a_1(1640)$ resonance.
            The blue solid and red solid curves represent the total and continuum spectral functions, respectively.
            The red dashed and dotted curves correspond to the resonant contributions from $a_1(1260)$ and $a_1(1640)$, respectively. 
            The black dashed-dotted lines denote the OPE spectral functions, serving as benchmarks for comparison. 
            The values of regularization parameter $\alpha$ are given at the top of each panel, and the extracted decay constants of pion and $a_1$ mesons are listed in the embedded boxes. 
            }
            \label{fig:result_under_regularization}
\end{figure}

We first solve the integral equation of Eq.~(\ref{eq:original_inverse_problem}) without regularization. 
The result is shown in Fig. \ref{fig:result_under_regularization}(a), which exhibits severe high-frequency oscillations and fails to produce any physically meaningful solution. 
This behavior confirms the ill-posed nature of the underlying inverse problem, and consequently demonstrates that regularization is essential for retrieving a stable and meaningful solution.
This numerical instability is consistent with the severe ill-conditioning of the matrix $K$, whose condition number is of order $10^{10}$.

For comparison, the black dash-dotted line represents the OPE spectrum, $\rho^{\rm OPE}(s)\equiv {\rm Im}\,\Pi_q^{\rm OPE}(s)/\pi$, as given in Eq.~(\ref{eq:perturbative_imaginary_Pi_q}).
However, at small $s$, the perturbative and OPE calculations are not reliable.
Therefore, we show the OPE spectrum only from $1$ GeV$^2$ upwards. 

We now apply Tikhonov regularization to the integral equation in Eq.~(\ref{eq:original_inverse_problem}), without incorporating the pion pole. 
The corresponding results are shown in Figs.~\ref{fig:result_under_regularization}(b) and (c) for the cases without and with boundary conditions, respectively. 
In contrast to the highly oscillatory non-regularized solution in Fig.~\ref{fig:result_under_regularization}(a), the regularized solutions are smooth and stable, demonstrating that Tikhonov regularization effectively mitigates the ill-posedness of the inversion.
These results are discussed below. 

The solution in Fig.~\ref{fig:result_under_regularization}(b) is obtained without any prior information or boundary conditions and is therefore relatively trivial.
In Fig.~\ref{fig:result_under_regularization}(c), we impose $\rho(t_{\rm min})=0$ at threshold and match the spectral function at $s=\Lambda$ to the imaginary part of the full OPE in Eq.~(\ref{eq:perturbative_imaginary_Pi_q}), as implied by quark-hadron duality at high energies.
The significant difference between Figs.~\ref{fig:result_under_regularization}(c) and~\ref{fig:result_under_regularization}(b) implies the importance of boundary conditions.
In particular, a bump-like structure emerges in Fig.~\ref{fig:result_under_regularization}(c), indicating the presence of hadronic states.
The bumps in the spectral function, potentially corresponding to resonances, can be attributed in part to the boundary conditions.
Since boundary conditions are a natural requirement of any physical problem,
the solution in Fig.~\ref{fig:result_under_regularization}(c) provides a more meaningful physical description.

The bumps shown in Fig.~\ref{fig:result_under_regularization}(c) are too broad and lack sufficient resolution.
As the pion pole is known a priori, we incorporate it into the spectrum as given in Eq.~(\ref{eq:spectral_decompose_N1}).
This procedure corresponds to the inverse problem in Eq.~(\ref{eq:final_equation}).
Taking only the pion mass as input, we extract the decay constant $f_\pi$.
The result is shown in Fig.~\ref{fig:result_under_regularization}(d).
For the continuum spectral function $\rho^c(s)$, the lower boundary is set at the $3\pi$ threshold, $s_{\rm th}=(3m_\pi)^2$, where $\rho(s_{\rm th})=0$.
The extracted value $f_\pi=122.3~\mathrm{MeV}$ comes reasonably close to the experimental value $f_\pi^{\rm exp}=130.4~\mathrm{MeV}$, demonstrating the effectiveness of the inverse problem approach.
The continuum spectral function still exhibits a broad bump, indicating additional hadronic structures beyond the pion pole.
However, the limited resolution prevents the extraction of detailed information about the underlying hadronic states.
The limited resolution stems from two main origins. First, input precision is insufficient, with OPE calculations performed up to next-to-leading order in $\alpha_s$ and dimension-six condensate contributions.
Second, Tikhonov regularization has limited resolving power for Breit-Wigner peaks, as discussed in Ref.~\cite{Xiong:2022}.
Further improvements to the reconstruction could be achieved, on the one hand, by refinements of the above two aspects, and on the other hand, by incorporating prior information on resonant states.

The axial-vector current $j_{\mu 5}$ couples to both pseudoscalar and axial-vector mesons.
Contributions from excited pseudoscalar mesons $\pi'$ to $\operatorname{Im}\Pi_q(s)$ are suppressed by the light current-quark masses $m_{u,d}$~\cite{Khodjamirian:2020btr,CLEO:1999rzk,Jiang:2015paa} and can thus be neglected.
By contrast, the axial-vector mesons do not suffer from such suppression.
We therefore examine the states $a_1(1260)$ and $a_1(1640)$ from the Particle Data Group (PDG)~\cite{ParticleDataGroup:2026mpi} in the following.
Unlike the pion pole, which is represented by a delta function, the $a_1$ resonances have finite widths and are described by relativistic Breit-Wigner forms.
The Breit-Wigner distribution for $a_1(1260)$ is given by
\begin{equation}
\mathbb{BW}\left(s,m_{a_1},\Gamma_{a_1}\right) = \frac{1}{\pi} \frac{m_{a_1} \Gamma_{a_1}(s)}{\left(s-m_{a_1}^2\right)^2 + m_{a_1}^2 \Gamma_{a_1}^2(s)}\,,
\end{equation}
where the energy-dependent width is defined as $\Gamma_{a_1}(s)=\Gamma_{a_1}h(s)/h(m_{a_1}^2)$, with $h(s)$ denoting the phase-space factor \cite{Bowler:1986exb,ParticleDataGroup:2026mpi}.
Note that $\lim_{\Gamma\to0}\mathbb{BW}\left(s,m,\Gamma\right)\to\delta(s-m^2)$.
The three-body decay $a_1\to3\pi$ involves a complicated phase-space structure, but it is dominated by the $a_1\to\rho\pi$ channel.
We therefore retain only the $\rho\pi$ contribution \cite{Bowler:1986exb}, for which
 \begin{equation}
  h(s)=\int_{4m_\pi^2}^{(\sqrt{s}-M_\pi)^2} \frac{d s'}{\pi} \frac{m_\rho \Gamma_{\rho}}{(s' -m_{\rho}^2)^2 + (m_{\rho} \Gamma_{\rho})^2} \frac{\sqrt{\lambda(s,s',m_\pi^2)}}{16 \pi s}\,,
 \end{equation}
with the Källén function  $\lambda(x, y, z) = x^2 + y^2 + z^2 - 2xy - 2yz - 2xz $.
Note that $h(s)$ vanishes at $s=(3m_\pi)^2$, so the boundary condition $\rho(s_{\rm th})=0$ is  naturally satisfied.
Finally, the spectral function including the pion and $a_1(1260)$ contributions is written as,
\begin{equation}{\label{eq:spectral_N2_a1-1260}}
    \rho(s)
       =f_{\pi}^2 \delta(s-m_{\pi}^2)
       +    f_{a_1(1260)}^2 \mathbb{BW}\left(s,m_{a_1(1260)},\Gamma_{a_1(1260)}\right) 
      + \rho^{c}(s) \,.
\end{equation}
Here, the prior information consists of $m_\pi$, $m_{a_1(1260)}$, and $\Gamma_{a_1(1260)}$, whereas $f_\pi$, $f_{a_1(1260)}$, and $\rho^c(s)$ are unknown and determined through the inverse problem approach.
The latest PDG average values are $m_{a_1(1260)}=1202\pm15~\mathrm{MeV}$ and $\Gamma_{a_1(1260)}=422\pm12~\mathrm{MeV}$.
These quantities previously had very large uncertainties, especially for the width.
Their precision has improved substantially, largely owing to the high-precision LHCb measurement of $D^{0} \rightarrow K^\mp \pi ^\pm \pi ^\pm \pi ^\mp$ \cite{LHCb:2017swu}.
Accordingly, we adopt these values in our analysis.
As the Breit-Wigner function is valid over the entire interval $[s_{\rm th}, \Lambda]$, the boundary condition at $s=\Lambda$ is given by $\rho(\Lambda)=\rho^{\rm OPE}(\Lambda)$, with $\rho(\Lambda)$ including both resonance and continuum contributions.

The final result is shown in Fig.~\ref{fig:result_under_regularization}(e). 
The red solid line represents the continuum density, the red dashed line corresponds to the $a_1(1260)$ resonance, and the blue solid line denotes the total spectral function.
Clearly, the $a_1(1260)$ peak indicates that this resonance contributes significantly to the spectral function.
More importantly, once the $a_1(1260)$ resonance is subtracted, the reconstructed continuum $\rho^c(s)$ is close to zero at low $s$ and begins to rise gradually above the $a_1(1260)$ resonance region. 
For $s\gtrsim4~\mathrm{GeV}^2$, the continuum and total spectral functions approach a plateau and agree closely with the OPE result.
This behavior not only demonstrates the sizable $a_1(1260)$ contribution, but also shows that the reconstructed continuum spectral function is physically reasonable and self-consistent, with no additional resonance observed.
Moreover, the result with $f_\pi=130.8$~MeV is closer to the experimental value of 130.4 MeV than that in Fig.~\ref{fig:result_under_regularization}(d).  

It remains unclear whether additional resonances contribute to the spectral function.
We therefore include the PDG-listed $a_1(1640)$ as further prior information, with the result shown in Fig.~\ref{fig:result_under_regularization}(f).
A clear $a_1(1640)$ peak appears, but the continuum near the upper endpoint shows poorer agreement with the OPE result than in Fig.~\ref{fig:result_under_regularization}(e).
As discussed in the preceding section, the transition between the perturbative and non-perturbative regimes is smooth, and the extracted results should remain insensitive to $\Lambda$ over a reasonable range.
Moreover, quark-hadron duality is expected to hold above $\Lambda$, so the reconstructed spectral function should approach the perturbative result near the upper end of the solution interval.
These features serve as an important criterion for assessing the reliability of the reconstructed solution.

Indeed, the continuum in Fig.~\ref{fig:result_under_regularization}(e) is nearly flat for $s\gtrsim4~\mathrm{GeV}^2$ and agrees closely with the perturbative result, whereas that in Fig.~\ref{fig:result_under_regularization}(f) exhibits fluctuations over $4\lesssim s\lesssim6~\mathrm{GeV}^2$.
We therefore favor the scenario in Fig.~\ref{fig:result_under_regularization}(e), which incorporates prior information only for the pion pole and the $a_1(1260)$ resonance.
At the current level of input precision, we expect the contributions from higher excited states to be absorbed in the continuum, as they cannot yet be resolved separately.
Resolving additional excited states will require more precise input and is left for future investigation.

So far, we have discussed the six cases shown in Fig.~\ref{fig:result_under_regularization} and selected the one in Fig.~\ref{fig:result_under_regularization}(e) as our final result.
Several remaining issues should now be addressed.

We first discuss the boundary condition at $s=\Lambda$. 
As seen in Fig.~\ref{fig:result_under_regularization}(c), bumps can arise from
boundary conditions. 
Besides, the solution in Fig.~\ref{fig:result_under_regularization}(f) is discarded because of a
bump near the right endpoint. 
For $s \in [4,6]~\mathrm{GeV}^2$, the spectral function first increases and then decreases, which is inconsistent with the OPE spectrum. 
This behavior appears to satisfy the boundary condition $\rho(\Lambda)=\rho^{\rm OPE}(\Lambda)$. 
However, $\rho^{\rm OPE}(\Lambda)$ is computed only up to a fixed order in $\alpha_s$ and the OPE, so it carries truncation uncertainties.
As a test, we removed the boundary condition at $s=\Lambda$ and examined how the solutions change.
For the solution in Fig.~\ref{fig:result_under_regularization}(e), the numerical results remain almost unchanged.
This indicates that the boundary condition is automatically satisfied and need not be imposed separately. 
For the solution in Fig.~\ref{fig:result_under_regularization}(f), removing the boundary condition eliminates the bump near $\Lambda$, but the spectral function becomes monotonically increasing and remains inconsistent with the flat OPE spectrum. 
Thus, when the solution itself is not correct, the discrepancy at the right endpoint cannot be resolved by either imposing or removing the boundary condition.
These observations suggest that the boundary condition at $s=\Lambda$ is not an independent requirement, but rather a consequence of a correct solution. 

In this work, the integral kernel of the dispersion relation is $1/[(s-q^2)(s-\sigma_0)]$, whereas that in Ref.~\cite{Xiong:2022} is $1/(s-q^2)$. 
In fact, the simpler form $1/(s-q^2)$ can equally be used here, since $s-\sigma_0$ is strictly positive.
The factor $1/(s-\sigma_0)$ can thus be absorbed into the redefinition of $\rho(s)/(s-\sigma_0)\to \tilde \rho(s)$ for the inverse problem in Eq.~(\ref{eq:original_inverse_problem}).
The integral equations associated with these kernels are equivalent.
To verify this equivalence numerically, we solve the inverse problem using both kernels.
The extracted values of $f_\pi$ are in excellent agreement, while those of $f_{a_1}$ differ by  only $0.1~\mathrm{MeV}$, confirming the numerical equivalence of the two formulations.

In principle, the spectral function $\rho(s)\propto\lvert\langle0|j_{\mu5}|n\rangle\rvert^2$, as given in Eq.~(\ref{eq:correlator-im}), should be non-negative.
Nevertheless, the reconstructed continuum in Fig.~\ref{fig:result_under_regularization}(e) exhibits slightly negative values in the low-$s$ region.
This occurs because non-negativity is not explicitly imposed as a constraint when solving the inverse problem in the current work.
For the present analysis, this violation is small and does not affect our discussion or conclusions.
Therefore, no dedicated non-negativity constraint was implemented in the numerical algorithm.
We expect this issue to be resolved in future work with more precise input. 

Finally, we discuss the incorporation of prior information about the pion and the $a_1$ mesons.
As discussed in \cite{Xiong:2022}, the solution of an inverse problem can generally be improved as more reliable prior information is incorporated.
The $L^2$ and $H^1$ Tikhonov regularizations penalize large values and gradients of the solution, respectively, and thus tend to over-smooth it and suppress peak structures.
This motivates the incorporation of prior information on the pion and the $a_1(1260)$. 
In principle, the peak structures can be reconstructed directly by a mathematical inversion method in the limit of vanishing input error, provided that convergence is guaranteed.
In practice, however, the input errors are finite, and we retain Tikhonov regularization together with the corresponding prior information in the present work. 
A direct reconstruction of these peak structures, without relying on such prior information, is left for future work.

\subsection{Uncertainty Analysis and Final Results}

In this section, we analyze the uncertainties in the reconstructed solution shown in Fig.~\ref{fig:result_under_regularization}(e).
In this case, the pion mass and the $a_1(1260)$ mass and width are taken as prior information, while the pion and $a_1(1260)$ decay constants and the continuum spectral function are determined by the inverse problem.
First, we examine the sensitivity of the reconstructed solution to variations of the relevant parameters, and find that plateaus exist for all of them.
Then, the input-induced and method-intrinsic uncertainties are analyzed. Importantly, the uncertainty arising from the approximation inherent in the method can be quantified, which demonstrates the reliability of the method.
Finally, the result is presented and compared with the experimental data to assess their agreement.

\paragraph{Stability Plateaus\\}

\begin{figure}
                \centering
                \subfigure[\rm $\alpha$-dependence]{\includegraphics[width=0.25\textwidth ]{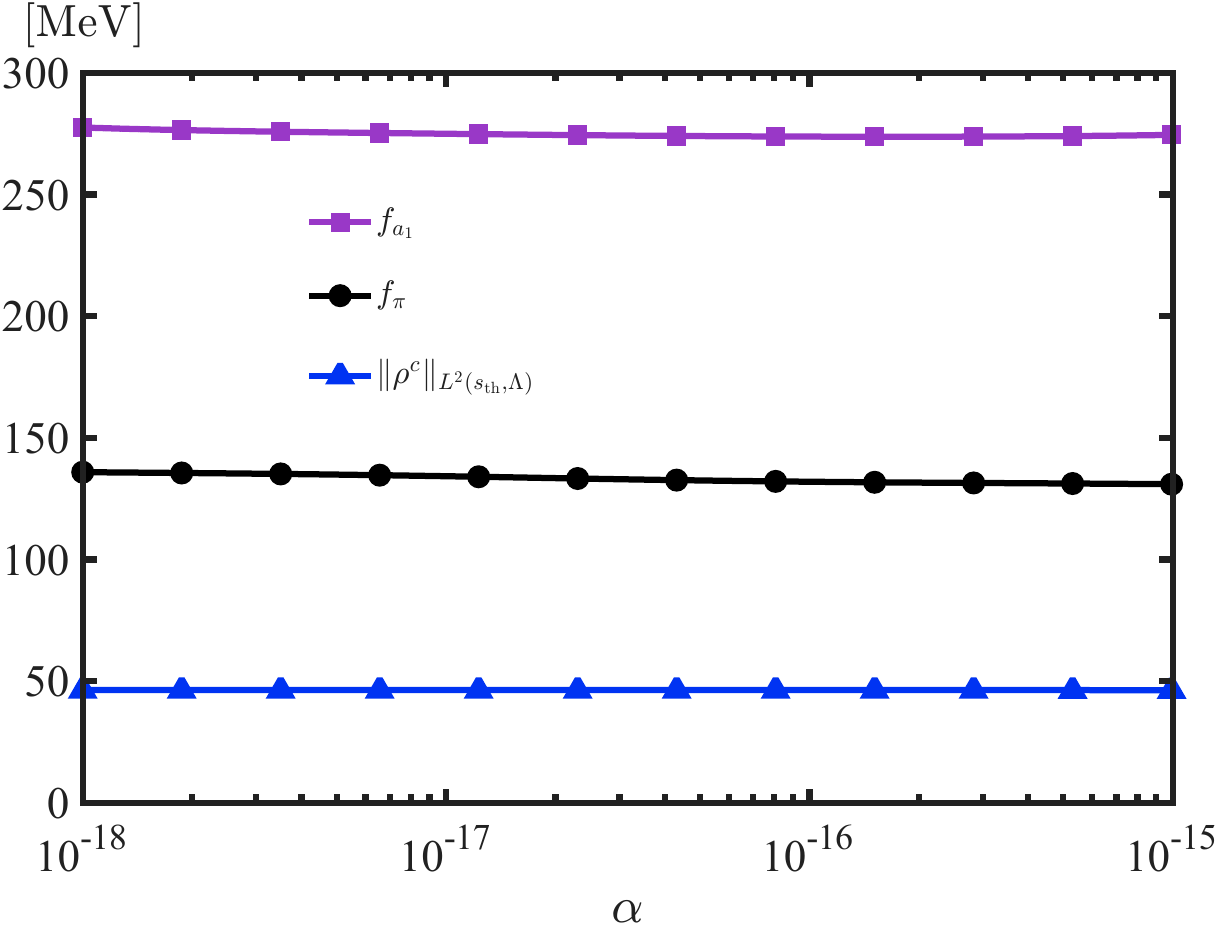}}
                 \hspace{5mm}
                \subfigure[\rm $\Lambda$-dependence]{\includegraphics[width=0.25\textwidth ]{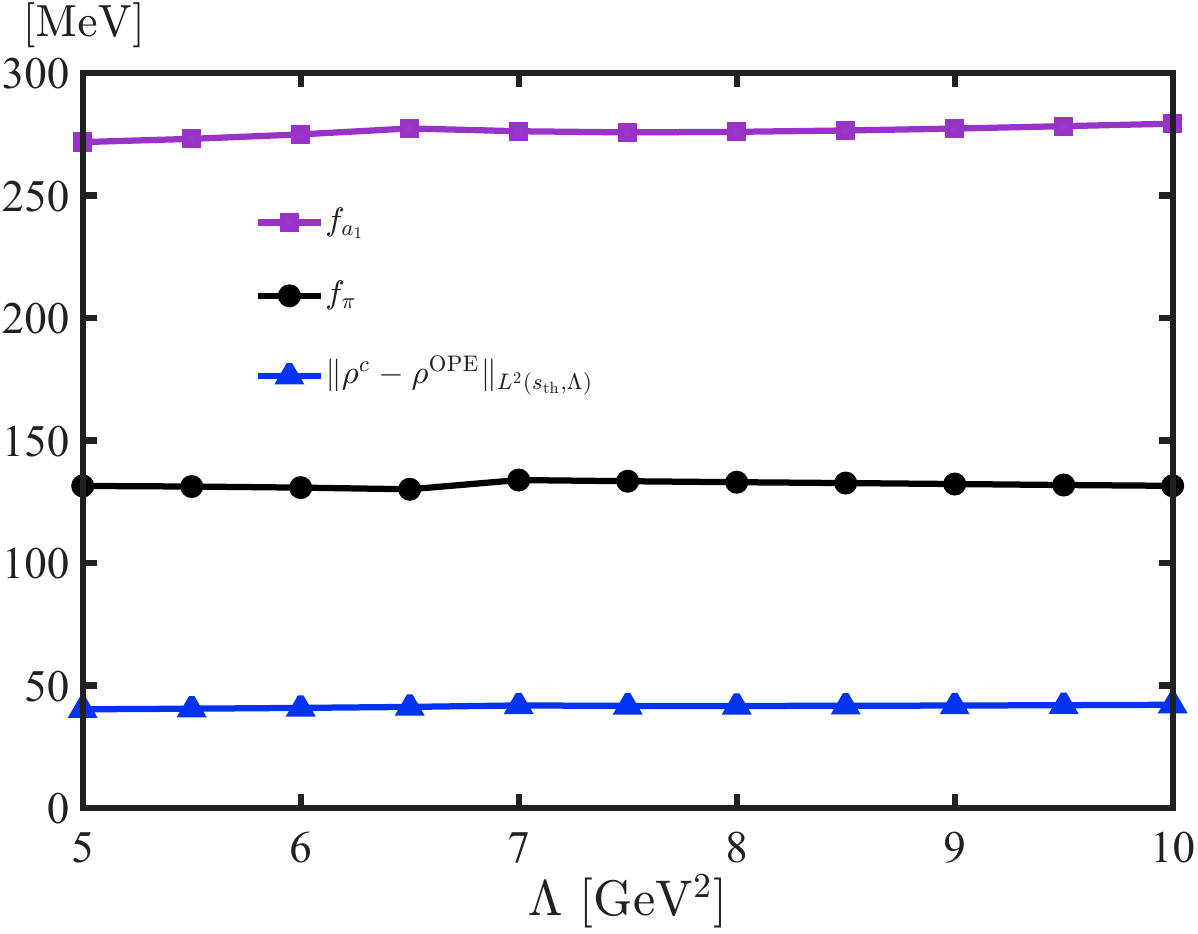}}
               \vspace{1mm}
               
                \subfigure[\rm $q_{\text{up}}^2$-dependence]{\includegraphics[width=0.25\textwidth ]{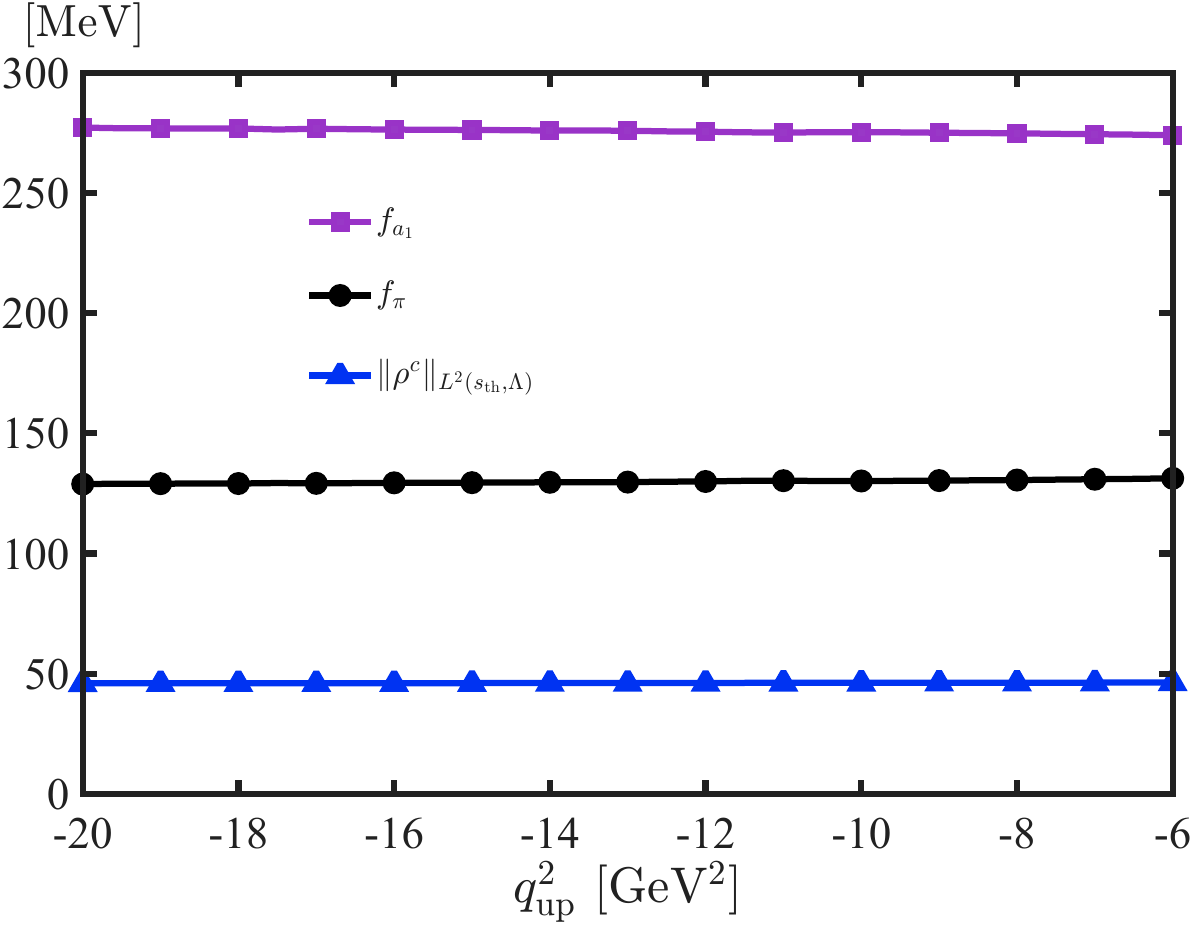}}
               \hspace{5mm}
               \subfigure[\rm $q_{\text{low}}^2$-dependence]{\includegraphics[width=0.26\textwidth ]{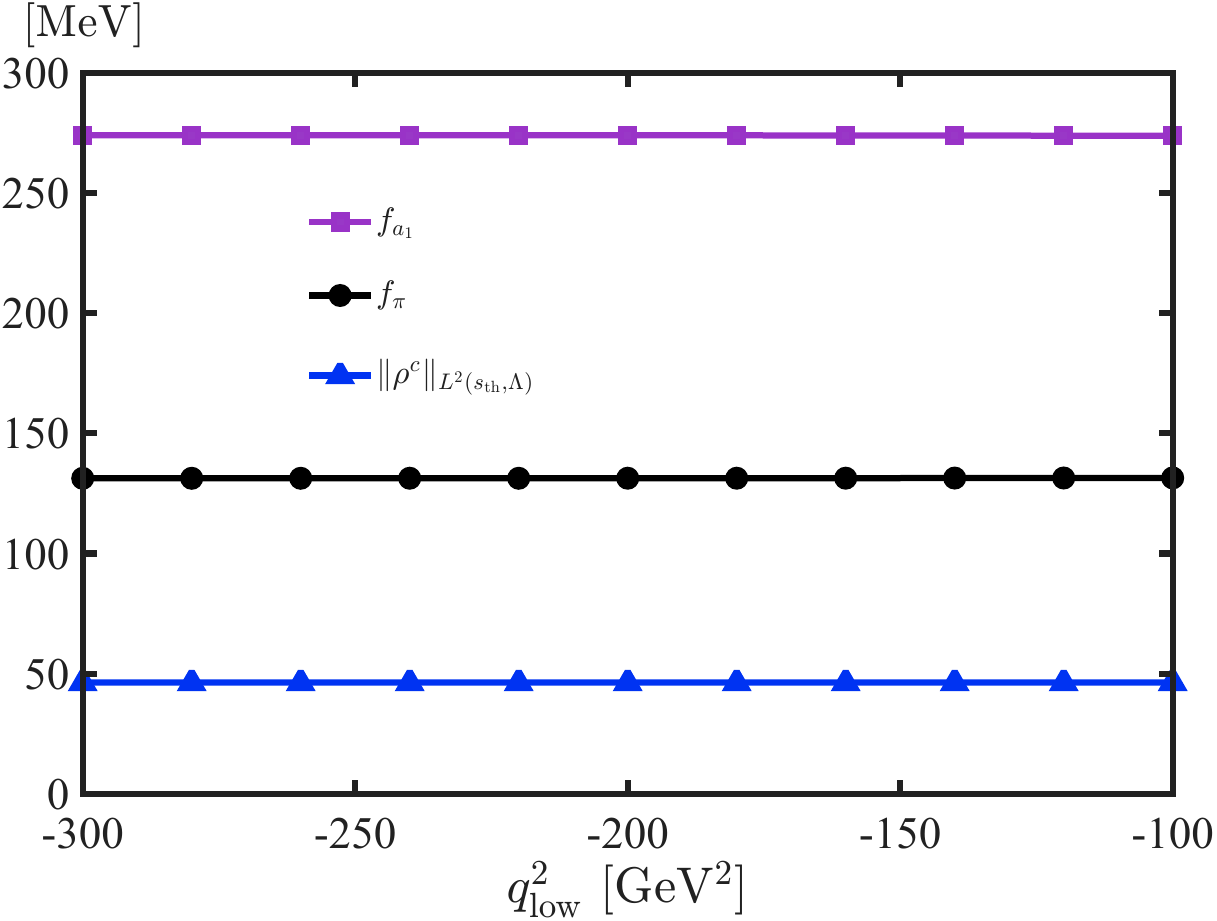}}
               \hspace{5mm}
                \subfigure[\rm $\sigma_0$-dependence]{\includegraphics[width=0.26\textwidth ]{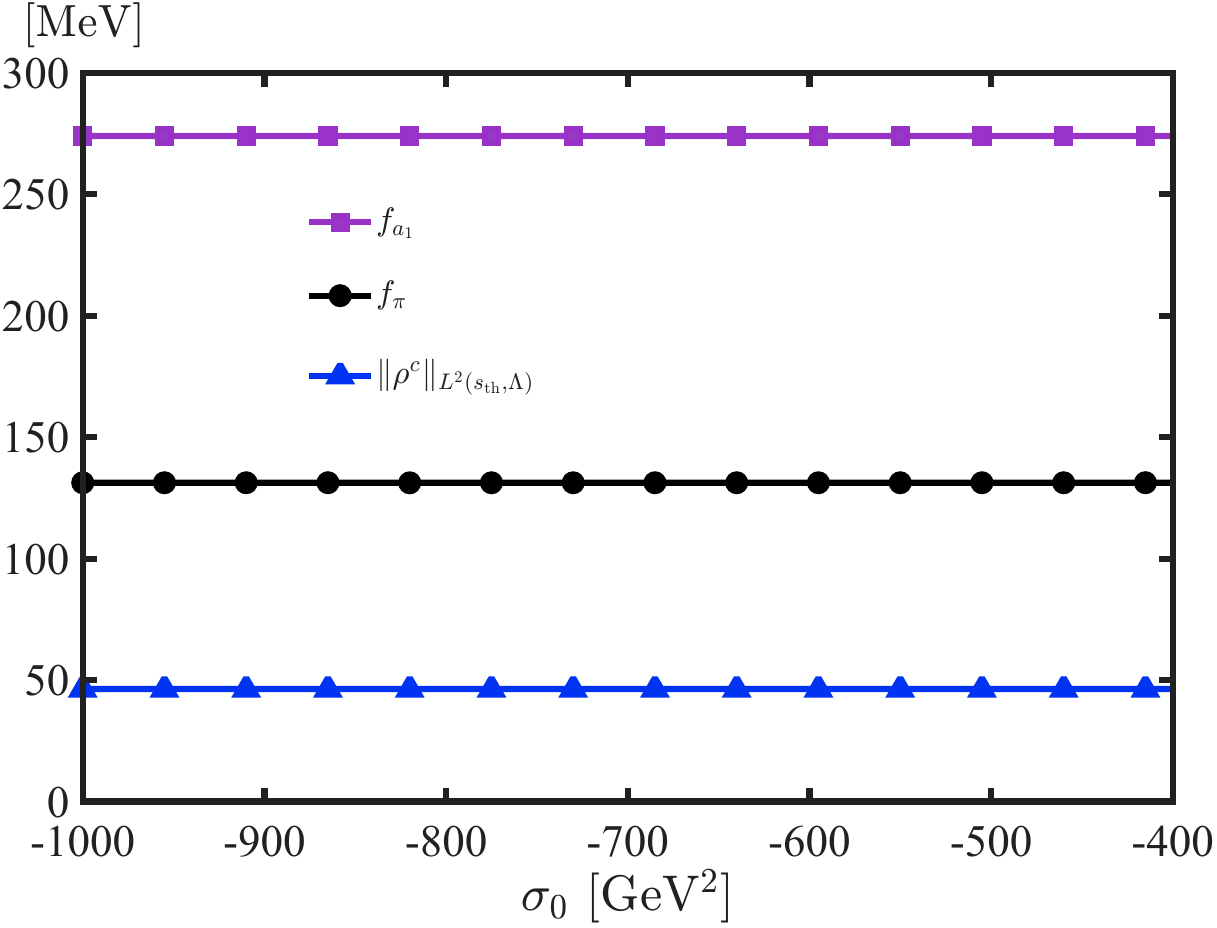}}
         \caption{Dependence of the reconstructed solution on the parameters of the inverse problem approach.
         }
             \label{fig:plateaus}
\end{figure}

In the present work, the parameters of the inverse problem approach include the regularization parameter $\alpha$, the perturbative/non-perturbative separation scale $\Lambda$, the subtraction point $\sigma_0$, and the upper and lower bounds of the input $q^2$ interval, $q_{\mathrm{up}}^2$ and $q_{\mathrm{low}}^2$.
The insensitivity to these parameters, manifested as stability plateaus, provides a key criterion for assessing the reliability of the inverse problem solution.
The dependence of the reconstructed solution on these parameters is shown in Fig.~\ref{fig:plateaus}.
The black and purple curves in the plots show the stability of the decay constants $f_\pi$ and $f_{a_1}$.
For the continuum spectral function, the blue curves in Fig.~\ref{fig:plateaus}(a) and (c)-(e) show $\|\rho^c\|_{L^2(s_{\rm th},\Lambda)}$.
Since this quantity increases almost linearly with $\Lambda$, we therefore use the difference $\|\rho^c-\rho^{\rm OPE}\|_{L^2(s_{\rm th},\Lambda)}$ to assess the stability of the continuum spectral function with respect to $\Lambda$.

The dependence of the results on the regularization parameter $\alpha$ provides a first test of the reliability of the inverse problem approach.
As shown in Fig.~\ref{fig:plateaus}(a), the stability window for $\alpha$ ranges from $10^{-18}$ to $10^{-15}$, covering about three orders of magnitude. 
Such a broad plateau demonstrates that the solution is robust against variations in $\alpha$, indicating that the result does not rely on fine-tuned regularization.
Furthermore, the quasi-optimality criterion selects $\alpha=10^{-16}$, which lies well within the plateau, confirming that this criterion reliably locates the stable region without manual tuning. 
Additionally, reducing $q^2_{\rm up}$ or $\sigma_0$ decreases the integral kernel $K$ and the input $g(q^2)$, thereby reducing the first term of $J_\alpha(f)$ in Eq.~\eqref{eq:Tikhonov_functional} and consequently shifting the broad plateau of $\alpha$ toward smaller values to maintain the same solution.

It is also important to test the sensitivity of the results to the separation
scale between perturbative and non-perturbative QCD, $\Lambda$. 
The corresponding results are shown in Fig.~\ref{fig:plateaus}(b), where the plateaus span $\Lambda \in [5,10]$~GeV$^2$. 
The wide stability window is consistent with the smooth transition between perturbative and non-perturbative QCD. 
$f_\pi$ is very flat, changing by less than 1~MeV, while $f_{a_1}$ shows slight variations of about 10~MeV.
For the continuum spectral function, $\|\rho^c-\rho^{\rm OPE}\|_{L^2(s_{\rm th},\Lambda)}$ changes by only 2~MeV.
This indicates that quark-hadron duality is a very good approximation over this range of $\Lambda$, and the extracted solution is very close to the OPE spectrum, as can be seen from Fig.~\ref{fig:result_under_regularization}(e).

Figs.~\ref{fig:plateaus}(c)-(e) show that the results are even more stable with respect to the parameters on the input side: the upper and lower bounds $q_{\rm up}^2$ and $q_{\rm low}^2$, and the subtraction point $\sigma_0$. 
All these parameters can vary over wide ranges. 
Indeed, they are all effective scales that do not possess a unique physical value.
Rather, they depend on the range over which the OPE calculations are valid, and hence on the precision of the input.
The broad plateaus indicate that, as long as the parameters lie within their ranges, the physical conclusions are independent of the specific choices of these parameters. 

Note that the inverse problem in Figs.~\ref{fig:plateaus}(b)-(e) varies point by point in principle, since the solution range is modified by $\Lambda$, or the inputs are changed by $q_{\rm low}^2$, $q_{\rm up}^2$ and $\sigma_0$.   
Accordingly, the $\alpha$ value differs at each point across these figures.
Nevertheless, all obtained $\alpha$ values lie within the stable window of $\alpha$. 
For this reason, the results do not exhibit significant variation and appear as plateaus. 

Finally, we emphasize that two criteria are crucial for assessing whether the inverse problem approach is valid for a given problem: the existence of stability plateaus in all parameters, and the closeness of the solution to the OPE spectrum near the right endpoint. 
If both criteria are satisfied, the results can be regarded as physically reliable, and the uncertainties can then be analyzed. 
If either is not satisfied, this may result from insufficient input precision or a regularization method that is not designed for the specific problem. 
In such a case, although a solution can still be obtained, its uncertainties may fail to cover the true physical values, making their discussion meaningless. 

As shown above, both criteria are satisfied in this work. 
The results exhibit broad plateaus for all parameters (Fig.~\ref{fig:plateaus}), and the solution approaches the OPE spectrum in the high-$s$ region (Fig.~\ref{fig:result_under_regularization}(e)).
This allows us to perform a detailed uncertainty analysis.

\paragraph{Input-Induced Uncertainties and Method-Intrinsic Uncertainties\\}

A sound theoretical method should allow its uncertainties to be estimated systematically and comprehensively, especially those arising from the assumptions or approximations inherent in the method itself.
In the present work, following Ref.~\cite{Xiong:2022}, the uncertainty is quantified by the difference between the regularized approximate solution $f_\alpha^\delta$ and the true solution $f$. 
It is divided into two contributions,
\begin{equation}\label{eq:uncertainty}
\left\|f_\alpha^\delta-f\right\|
\leq
\delta\left\|R_\alpha\right\|
+
\left\|R_\alpha Kf-f\right\|.
\end{equation}
The first term quantifies the uncertainty originating from the input error $\delta$.
The second term describes the approximation uncertainty arising from replacing the exact inverse $K^{-1}$ with the regularized operator $R_\alpha$ at finite $\alpha$, which manifests as a systematic bias in the reconstructed solution. 

Accordingly, the uncertainties of the solutions can be categorized into two components: input-induced uncertainty and method‑intrinsic uncertainty. 
The input-induced uncertainty includes errors in input parameters such as vacuum condensates and the mass and width of $a_1(1260)$, as well as truncation uncertainties arising from the $\alpha_s$ expansion and the OPE. 
These uncertainties stem purely from external inputs and are not directly related to the inverse problem solution procedure. 
The second component, method-intrinsic uncertainty, originates within the inverse problem approach. It contains two parts: the approximation uncertainty induced by finite regularization, and contributions from variations of framework-specific parameters, including $\Lambda$, $q_{\rm low}^2$, $q_{\rm up}^2$, and $\sigma_0$, within their stability plateaus. 

It is useful to contrast this classification with the conventional statistical/systematic categorization of uncertainties used in experimental or lattice-QCD analyses. 
There, statistical uncertainties stem from measured or lattice-simulated data, while systematic uncertainties originate from analysis-specific procedures.
In that conventional classification, the distinction is essentially based on whether the uncertainty is stochastic or bias-dominated. 
Our classification, by contrast, is based on the source of the uncertainties: input-induced uncertainty comes from external inputs, while method-intrinsic uncertainty originates within the inverse-problem approach itself. 
We therefore do not map our two components onto statistical and systematic uncertainties.
Instead, we distinguish them based on whether they come from outside the method or from the method itself.

\begin{table*}[t]
  \centering
  \caption{input-induced uncertainties and method-intrinsic uncertainties of $f_\pi$ and $f_{a_1(1260)}$. }
  \label{tab:uncertainties}
  \small
  \setlength{\tabcolsep}{7pt}
  \begin{tabular}{lllrr}
    \hline \hline
     & Source & Prescription
      & $\Delta f_\pi$ [MeV] & $\Delta f_{a_1}$ [MeV] \\
    \toprule
   \multirow{11}{1cm}{\textit{Input Errors}} 
       & $m_{u(d)}$ & $2.16\pm0.04\,(4.70\pm 0.04)\,\rm[MeV]$
      & 0.01 &  0.01 \\
       & $\myconds{\bar{q}q}$ & $-(0.272 \pm 0.005)^3\,[\mathrm{GeV}]^3$
      & 0.57 &  1.03 \\
      & $\langle\alpha_s GG\rangle$ & $0.07\pm 0.02\, [\rm GeV^4]$
      & 3.51 &  4.57 \\
      & $M_{a_1(1260)}$ & $1202\pm15\,[\mathrm{MeV}]$
      & 0.78 & 3.52 \\
      & $\Gamma_{a_1(1260)}$ &  $422\pm12\,[\mathrm{MeV}]$
      & 0.08 &   2.05 \\
      & $\mu$ & $[1,3]\, \mathrm{GeV}$
      &  0.77 &  2.41 \\
      & OPE truncation & dim-6 operators &   6.73 & 11.30\\
    \cmidrule(lr){2-5}
      &  Total input-induced uncertainty & &\textbf{7.69} &  \textbf{13.12} \\
    \midrule
    \multirow{6}{1cm}{\textit{Method Errors}} 
      & $\alpha$ &  $4\times10^{-16}$
      &  0.34&  0.40\\
      & $\Lambda$ & $[5,10]\, \mathrm{GeV}^2$
      & 2.11 &   3.77 \\
      & $q^2_{\mathrm{up}}$ & $[-20,-6]\, \mathrm{GeV}^2$
      &  1.19 &  1.49 \\
      & $q^2_{\mathrm{low}}$ & $[-300,-100]\, \mathrm{GeV}^2$
      &   0.08 &   0.10 \\
      & $\sigma_0$ & $[-1000,-400]\, \mathrm{GeV}^2$
      & 0.02 &   0.03 \\
    \cmidrule(lr){2-5}
      &  Total method-intrinsic uncertainty & & \textbf{2.45} & \textbf{4.11}\\ \bottomrule
    \multicolumn{3}{l}{Total uncertainty}
      &  \textbf{8.07} &   \textbf{13.74} \\
    \hline\hline
  \end{tabular}
\end{table*}

We now turn to the uncertainty analysis. 
The first category is input-induced uncertainty.
One source is the input parameters, including the vacuum condensates and the mass and width of the $a_1(1260)$, whose values are listed in the third column of Table~\ref{tab:uncertainties}. 
We sample each parameter over its uncertainty range and thus obtain the uncertainties of $f_\pi$, $f_{a_1}$, and the continuum spectrum. 
The resulting uncertainties of $f_\pi$ and $f_{a_1}$ are shown in the last two columns of Table~\ref{tab:uncertainties}.
Among these, the gluon condensate $\langle\alpha_s GG\rangle$ is the dominant source, contributing $3.51$~MeV to $f_\pi$ and $4.57$~MeV for $f_{a_1}$.
For $f_{a_1}$, the mass and width of the $a_1(1260)$ also contribute significantly.
The remaining condensates contribute at or below the few-MeV level. 

The second important source of input-induced uncertainty is the truncation of the $\alpha_s$ expansion and the OPE. 
The $\alpha_s$ truncation uncertainty is conventionally estimated by varying the renormalization scale $\mu$. 
In our calculation, the $\alpha_s$ terms are evaluated by varying $\mu$  from 1 to 3 GeV, taking the central value at $\mu=2$ GeV. 
This variation induces an uncertainty of 0.77~MeV in $f_\pi$ and 2.41~MeV in $f_{a_1}$. 
The OPE truncation uncertainty is estimated from the contribution of the highest-dimensional operators retained. 
Since $Q^2$ is sufficiently large in our kinematic region, the contributions of higher-dimensional operators are smaller than those of lower-dimensional ones, as explicitly demonstrated in Sec.~\ref{sec:condensates}.
In our analysis, the contribution of dimension-6 operators serves as a reasonable upper bound on the contributions of all higher-order terms.
Therefore, we quantify this uncertainty by the difference between the results with and without the contribution of dimension-6 operators.
This difference contributes a sizable uncertainty, $6.73$ MeV for $f_\pi$ and $11.30$ MeV for $f_{a_1}$, which is the dominant source of input-induced uncertainty.
Combining all individual contributions in quadrature, we obtain the total input-induced uncertainties of $7.69$~MeV for $f_\pi$ and $13.12$~MeV for $f_{a_1}$, summarized in the upper part of Table~\ref{tab:uncertainties}.

The method-intrinsic uncertainties of $f_\pi$ and $f_{a_1(1260)}$ are listed in the lower part of Table~\ref{tab:uncertainties}.
The first source of the method-intrinsic uncertainty originates from the approximation associated with the finite regularization parameter $\alpha$. 
The original problem corresponds to $\alpha\to 0$, but this limit is ill-posed and cannot be solved directly. 
The regularization method introduces a finite parameter $\alpha$ to convert the original ill-posed problem into an approximately well-posed one. 
In this way, the problem can be solved, but only at the cost of introducing an approximation. 
It is therefore necessary to quantify the uncertainty resulting from this approximation. 
As given in Eq.~(\ref{eq:uncertainty}), this approximation error can be estimated by $\| R_\alpha K f- f \|$. 
In practice, we estimate this uncertainty by replacing the unknown true solution $f$ with the numerical solution $f_\alpha^\delta$ obtained from Eq.~(\ref{eq:f_alpha^delta}).
As shown in Table~\ref{tab:uncertainties}, this uncertainty amounts to only 0.34 MeV for $f_\pi$ and 0.40 MeV for $f_{a_1}$, which is much smaller than the other sources of uncertainty. 
This indicates that the approximation-induced uncertainty of the inverse problem is both controllable and small in this work. 
We note that the estimate of the approximation uncertainty relies on the satisfaction of two criteria: the existence of stability plateaus and the consistency between the solution and the OPE spectrum at large $s$.

A further method-intrinsic uncertainty stems from the variation of the method parameters $\Lambda$, $q^2_{\rm low}$, $q_{\rm up}^2$, and $\sigma_0$ within their respective stability plateaus. 
Among them, the variation of $\Lambda$ gives the largest contribution, shifting $f_\pi$ by $2.11$~MeV and $f_{a_1}$ by $3.77$~MeV. 
The remaining parameters, $q^2_{\mathrm{up}}$, $q^2_{\mathrm{low}}$, and $\sigma_0$, play only a minor role, each contributing less than $1$~MeV. 
Combining these contributions with the above approximation-induced uncertainty in quadrature, the total method-intrinsic uncertainty amounts to $2.45$~MeV for $f_\pi$ and $4.11$~MeV for $f_{a_1}$.

The sources of uncertainty and their results are listed in Table~\ref{tab:uncertainties}.
Combining the input-induced and method-intrinsic uncertainties, we obtain the total uncertainties $\Delta f_\pi=8.07$~MeV and $\Delta f_{a_1}=13.73$~MeV. 
Moreover, the method-intrinsic uncertainties are smaller than the input-induced ones for both $f_\pi$ and $f_{a_1}$, indicating that the inverse problem approach and the regularization method adopted here are safe and reliable. 
Reducing the input uncertainty mainly requires more precise determination of the vacuum condensates and the calculation of higher-dimensional operators.
In general, improvements in all aspects of the input calculations are needed, including high-order loop corrections and more precise measurements of the resonance masses and widths.

\paragraph{Final Results and Comparison with Experiments and Other Methods\\}

\begin{figure}
    \centering
    \includegraphics[width=0.6\textwidth]{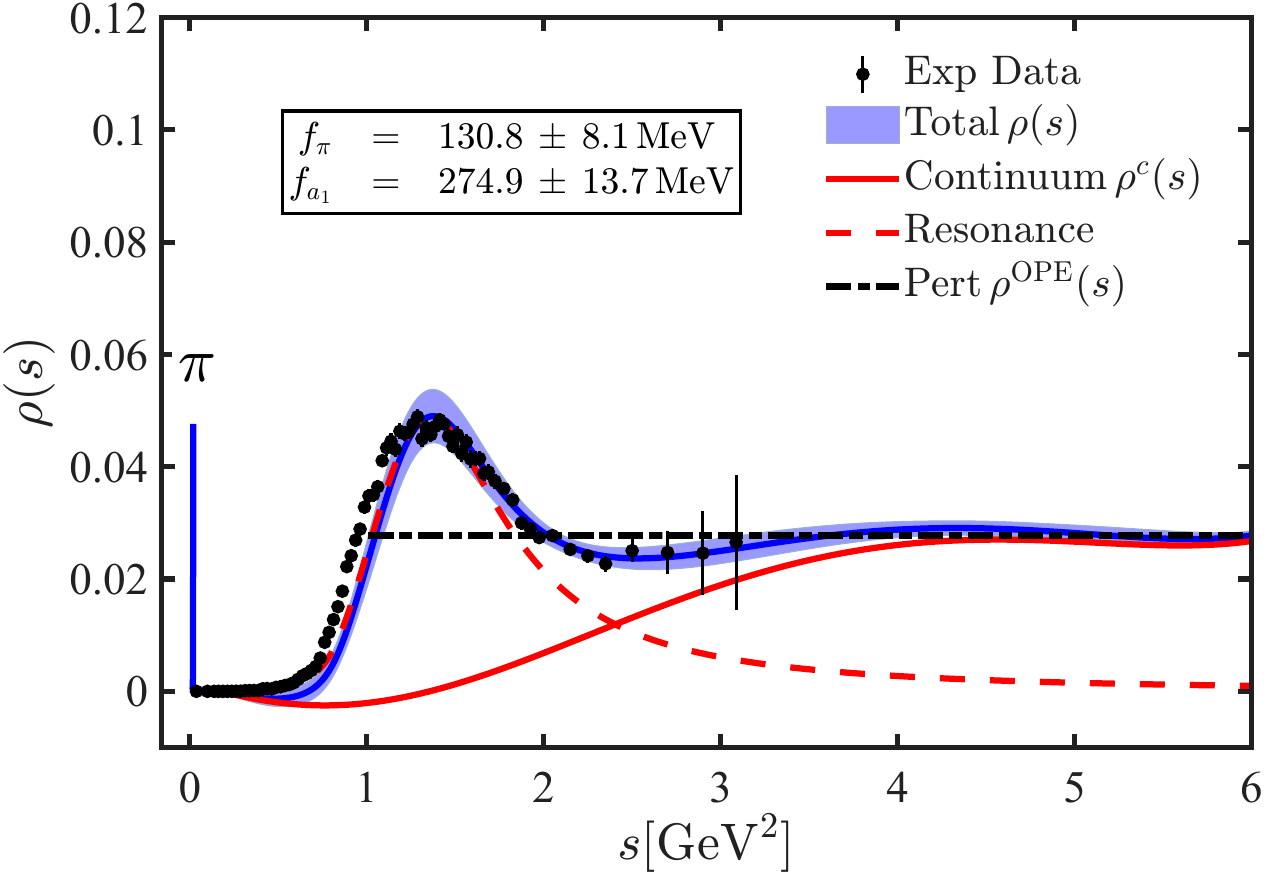}        
    \caption{Final results of the decay constants and spectral function obtained from the inverse problem approach. The continuum contribution $\rho^c(s)$ (red solid line), the $a_1(1260)$ resonance contribution (red dashed line), and the total spectral function $\rho(s)$ (blue solid line, with shaded uncertainty band) are shown, while the pion and $a_1(1260)$ decay constants are given in the inset. Black dots with error bars are experimental data.
     }
            \label{fig:final_result}
\end{figure}

The final results and their estimated uncertainties are presented in Fig.~\ref{fig:final_result}. The decay constants of the pion and $a_1(1260)$ are provided in the inset block. The continuum contribution $\rho^c(s)$ is shown as the red solid line, while the $a_1(1260)$ resonance contribution is depicted by the red dashed line. The total spectral function $\rho(s)$ is represented by the blue solid line, with the shaded band indicating the corresponding uncertainty. The black dots with error bars correspond to the experimental data.

We first compare $f_\pi$ with the experimental value as well as the determinations from lattice QCD and QCD sum rules in Table~\ref{tab:Results_Decay_constants}.
Our result, $f_\pi = 130.8\pm 8.1$~MeV, agrees with the experimental value $130.57\pm0.14$~MeV within uncertainties \cite{ParticleDataGroup:2026mpi}.
It is also consistent with the lattice QCD determination $130.20 \pm 0.80$~MeV~\cite{FlavourLatticeAveragingGroupFLAG:2024oxs} and the QCD sum rule results $127\pm15$~MeV~\cite{Shifman:1978by,Colangelo:2000dp,Khodjamirian:2020btr} and $135\pm1$~MeV~\cite{Jiang:2015paa}.
At the present level of input precision, our uncertainty remains larger than those of the experiment and lattice QCD.
Accordingly, the precision of the extracted $f_\pi$ alone does not yet reveal the full capability of the inverse problem approach.
Instead, the key strength of this approach lies in its simultaneous access to excited states and the continuum spectral function.

We next turn to the full continuum spectral function, which can be directly compared with experimental data.
The black points with error bars in Fig.~\ref{fig:final_result} represent the experimental spectral data extracted from $\tau$-decay measurements by the ALEPH Collaboration \cite{ALEPH:2005qgp,Pich:2013lsa,Davier:2013sfa,ALEPH_spectral_data}.
These data correspond to the axial-vector channel and include the sum over all relevant hadronic final states, such as $3\pi$, $5\pi$, and $\pi K K$.
As illustrated in Fig.~\ref{fig:final_result}, our numerical result shows excellent agreement with the experimental data.
This consistency constitutes one of the most compelling validations of the inverse problem approach.

From the threshold $s_{\rm th}=(3m_\pi)^2\simeq 0.17$~GeV$^2$ up to $s=m_\tau^2\simeq 3.16$~GeV$^2$, the reconstructed spectrum agrees with the experimental data across almost the entire range.
In the region $(3m_\pi)^2\leqslant s\lesssim2~\mathrm{GeV}^2$, the spectrum is dominated by the $a_1(1260)$ resonance.
While its peak position and width are fixed by prior information from experiments, the amplitude of the resonance is governed by the extracted $f_{a_1}$.
The observed agreement with experimental data thus demonstrates that $f_{a_1}$ is well constrained.
A small shift of the spectrum relative to the experiment appears at low s, arising because the input width of $a_1(1260)$ incorporates only the $\rho\pi$ contribution, whereas the physical experimental line shape involves more intricate decay dynamics \cite{GomezDumm:2003ku}.
Moreover, the PDG input is based primarily on measurements of $D\to K\pi\pi\pi$ decay~\cite{LHCb:2017swu}, which may introduce an additional error.

\begin{table}
    \centering
    \caption{Decay constants of pion and $a_1(1260)$. Our results are compared with experimental data, as well as determinations from lattice QCD and the QCD sum rules. For QCD sum-rule calculations, the correlation functions $\Pi_q$ and $\Pi_g$ are adopted respectively. }
    \label{tab:Results_Decay_constants}
    \begin{tabular}{@{}l|c|c}
    \hline 
    \hline 			
    Method & $f_{\pi}$ [MeV] & $f_{a_1(1260)}$ [MeV] \\
    \hline
    This work & 
    $130.8 \pm 8.1$& 
    $274.9 \pm 13.7 $\\
    Experiment & $130.57\pm0.14$ \cite{ParticleDataGroup:2026mpi} &  \\
    Lattice QCD & 
    $130.20 \pm 0.80$~\cite{FlavourLatticeAveragingGroupFLAG:2024oxs} & 
    $240.0 \pm 28.5$~\cite{Wingate:1995hy} \\
    QCDSR~($\Pi_q$) & 
    $127\pm15$~\cite{Colangelo:2000dp}, $135 \pm 1$~\cite{Jiang:2015paa} & 
    $151 \pm 20$~\cite{Jiang:2015paa} \\
     QCDSR~($\Pi_{g}$) & 
    & 
    $238 \pm 10$~\cite{Yang:2007zt}\\
    \hline 
    \hline 		   
    \end{tabular}
    {\footnotesize 
    \begin{itemize}
        \item[Note:] 
        In lattice studies \cite{Wingate:1995hy}, the reported values $f_{a_1(1260)}=0.30\pm0.03\, [\rm GeV^2]$ carries units of GeV$^2$, corresponding to the product $f_{a_1}m_{a_1}$ in our notation.
        To facilitate a direct comparison, we convert their results by dividing by the $a_1$ mass quoted in their publications.
    \end{itemize}}
\end{table}

At higher energies, the agreement with experiment over $2~\mathrm{GeV}^2\lesssim s\lesssim3~\mathrm{GeV}^2$ validates the accuracy of the reconstructed continuum spectral function $\rho^c(s)$.
Besides, despite substantial experimental uncertainties near $s=3~\mathrm{GeV}^2$, our result remains in close agreement with the experimental central values.
For $4~\mathrm{GeV}^2\lesssim s\lesssim6~\mathrm{GeV}^2$, the reconstructed spectrum $\rho(s)$ closely matches the OPE spectral function $\rho^{\rm OPE}(s)$.
This observation provides further validation of the theoretical framework and supports quark-hadron duality within this kinematic window.

Finally, we discuss the decay constant of the $a_1(1260)$.
Table~\ref{tab:Results_Decay_constants} compares our result with determinations from lattice QCD and QCD sum rules.
Our value, $274.9 \pm 13.7$~MeV, is larger than those obtained from both methods, $240.0\pm28.5$~MeV from lattice QCD~\cite{Wingate:1995hy}, $151\pm20$~MeV~\cite{Jiang:2015paa} or $238\pm10$~MeV~\cite{Yang:2007zt} from QCD sum rules.
No direct experimental value exists for $f_{a_1}$, yet its magnitude may be inferred from the spectral data shown in Fig.~\ref{fig:final_result}.
With our extracted $f_{a_1}$, the resulting spectral function agrees well with the experimental data, therefore providing an independent validation of our result.
The lattice QCD determination was reported in 1995~\cite{Wingate:1995hy}, and its uncertainty should be interpreted in the context of the computational limitations of that era.
Two QCD sum-rule determinations of $f_{a_1}$ can be found in Refs.~\cite{Yang:2007zt,Jiang:2015paa}.
Both works adopt the zero-width approximation for the $a_1(1260)$, describing its spectral contribution via a delta function.
In Ref.~\cite{Yang:2007zt}, $f_{a_1}$ is calculated from the correlation function $\Pi_g$ given in Eq.~\eqref{eq:correlator}.
As the $a_1(1260)$ is the lowest-lying state in this setup, this extraction is relatively clean and less prone to systematic bias.
By contrast, Ref.~\cite{Jiang:2015paa} employs the correlation function $\Pi_q$, which receives contributions from both the pion and $a_1(1260)$, complicating the simultaneous extraction of their parameters.
The resulting value from \cite{Jiang:2015paa} is too small to accommodate the experimental spectral data. 
This difficulty in simultaneously describing multiple states highlights a key advantage of the inverse-problem method.

\section{Physical Discussions}\label{sec:physical-discussions}
In this section, we discuss the physical implications of the results obtained from the inverse problem approach.

\subsection{Convergence of Solutions}\label{sec:convergence}

A crucial property of the inverse problem approach is that the approximate solution converges to the true solution as the input error $\delta\to 0$.
A rigorous mathematical proof of this convergence was given in Ref.~\cite{Xiong:2022}.
In this subsection, we examine this convergence property for the specific physical problem considered here.

\begin{figure}
                \centering
               \subfigure[\rm pure LO]{\includegraphics[width=0.3\textwidth ]{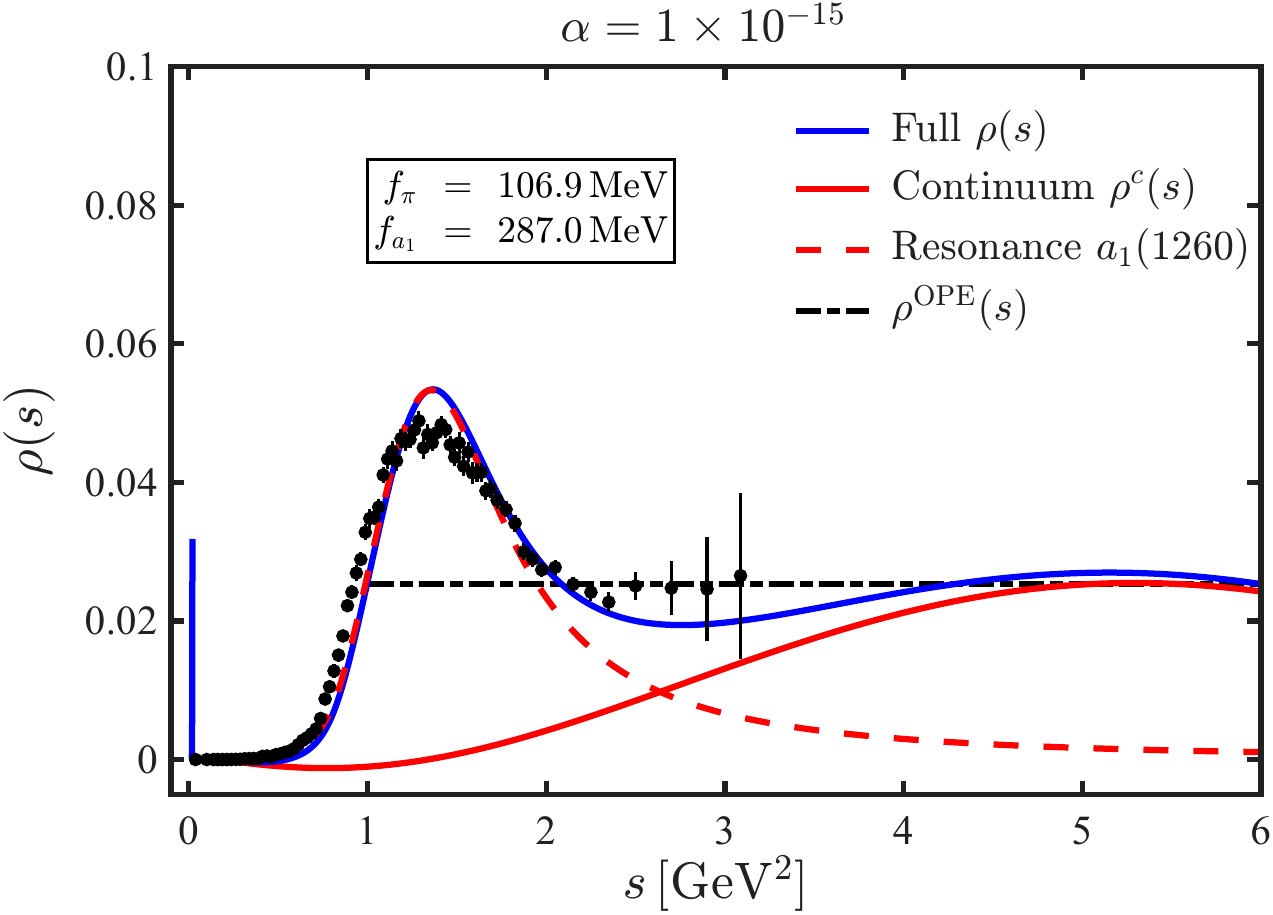}}
               \hspace{5mm}
               \subfigure[\rm LO+Condensates]{\includegraphics[width=0.3\textwidth ]{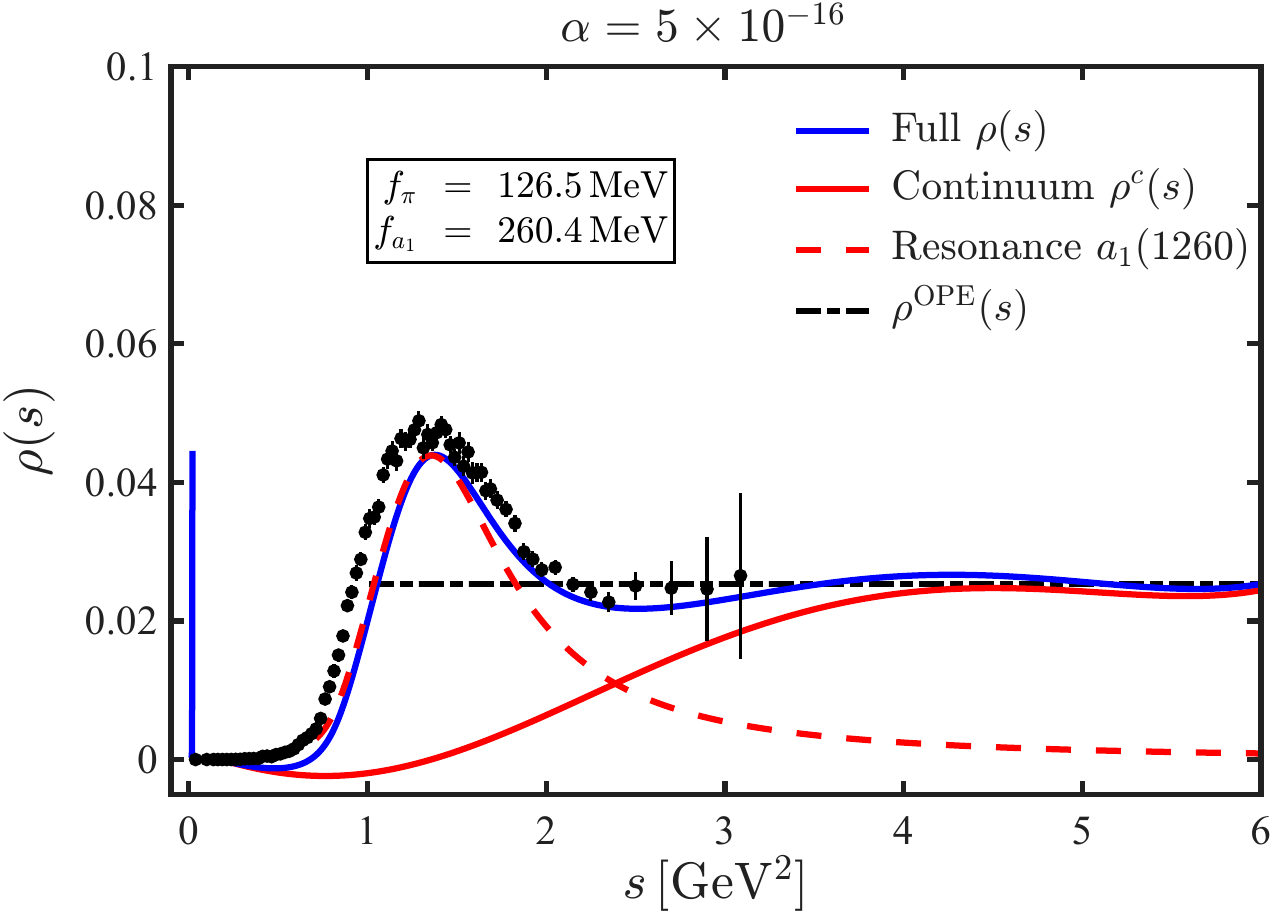}}
               \hspace{5mm}
               \subfigure[\rm LO + NLO + Condensates]{\includegraphics[width=0.3\textwidth ]{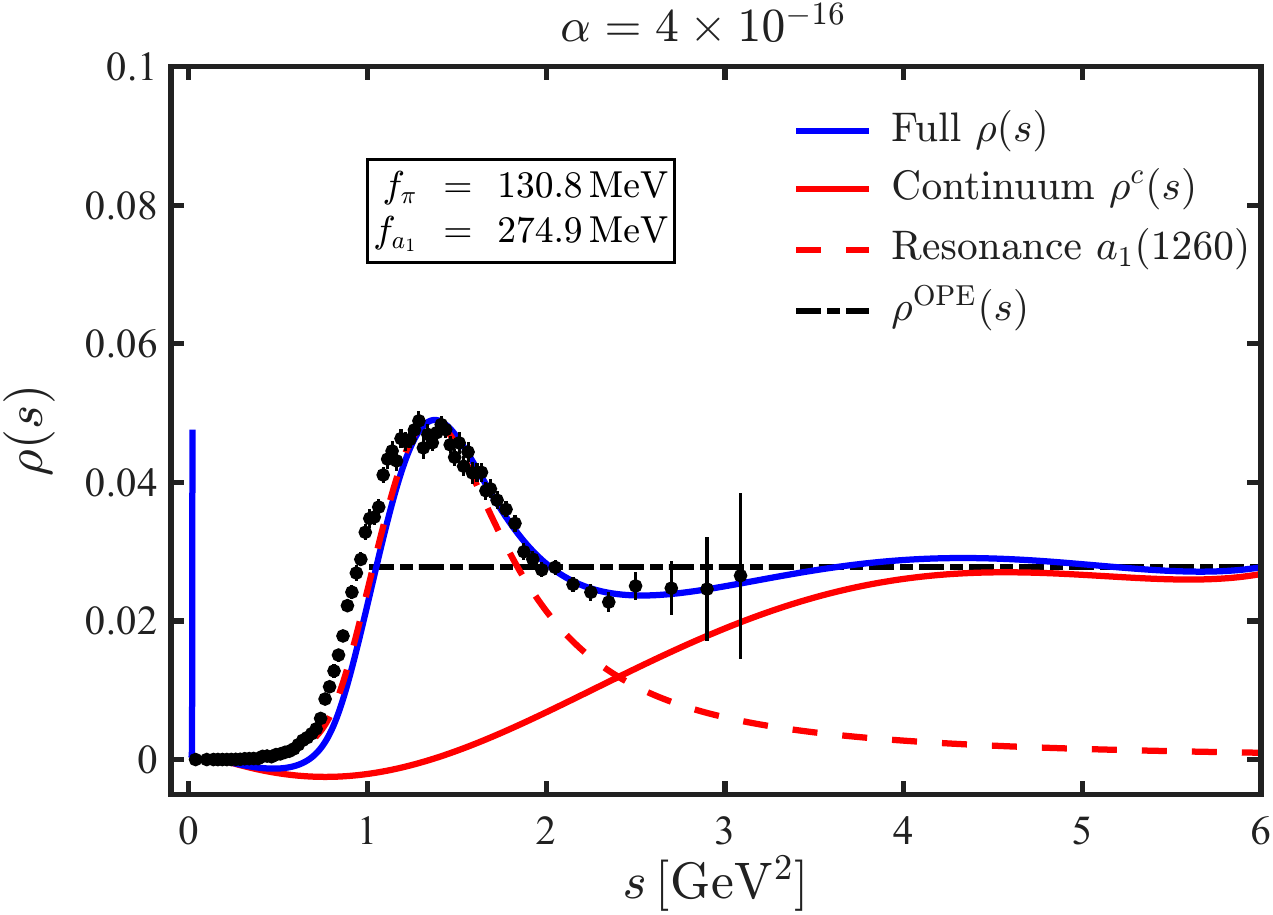}}
               \vspace{2mm}
               
               \subfigure[\rm pure NLO]{\includegraphics[width=0.3\textwidth ]{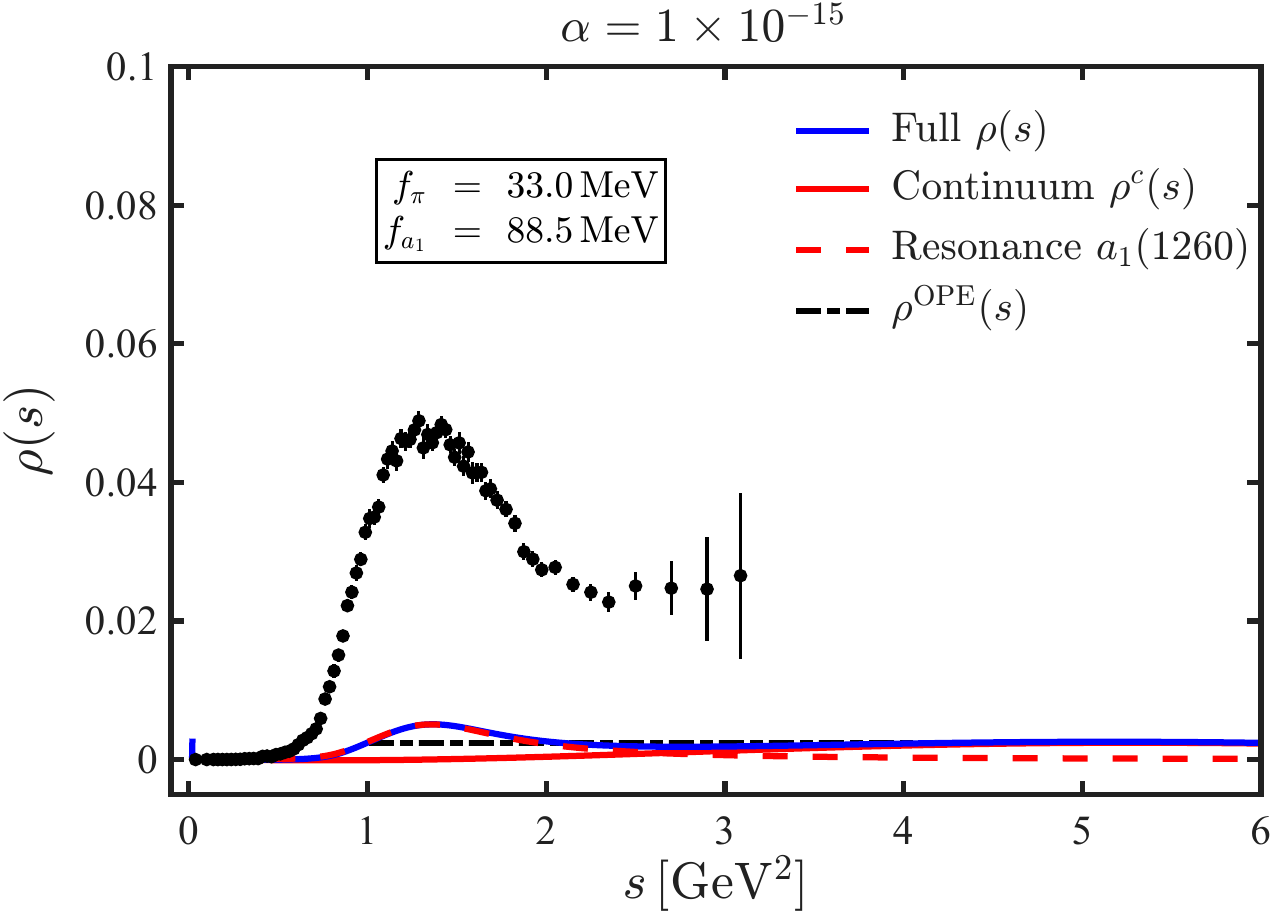}}

            \caption{Convergence of the solutions obtained with inputs from (a) pure LO perturbative contributions, (b) LO perturbative terms plus vacuum condensates, and (c) full  LO and NLO perturbative terms together with vacuum condensates. 
            For comparison, panel (d) shows the result with the pure NLO perturbative term as input. }
            \label{fig:Convergence}
\end{figure}

We consider four levels of theoretical input given in Eq.~\eqref{eq:OPE_calculation}: (a) only the pure leading-order (LO) perturbative term; (b) the LO contribution together with the vacuum condensates; (c) the full contribution from the LO and next-to-leading-order (NLO) perturbative terms and the vacuum condensates; and (d) only the pure NLO contributions for comparison.
The corresponding results are shown in the four panels of Fig.~\ref{fig:Convergence}.
With the successive inclusion of more contributions, the extracted $f_\pi$ changes from (a) 106.9~MeV for the pure LO input, to (b) 126.5~MeV for the LO+condensates input, and finally to (c) 130.8~MeV for the full input, the last being close to the experimental value of $130.57\pm0.14$~MeV. 
The result thus approaches the experimental value as the theoretical input is improved, exhibiting clear convergence.
Moreover, from Fig.~\ref{fig:Convergence}(a) to (c), the extracted spectral function shows better agreement with the experimental data at low $s$ and with the OPE spectrum at large $s$.
This example demonstrates that the predictive performance of the inverse problem approach can be systematically improved: the more precise the input, the more accurate the reconstruction.

Fig.~\ref{fig:Convergence}(d) shows the result obtained with only the NLO perturbative contribution as input.
The reconstruction is unsatisfactory: $33.0$~MeV for $f_\pi$ and $88.5$~MeV for $f_{a_1}$, both far from the experimental values, and the resulting spectrum is likewise inconsistent with the data.
This is physically expected since the NLO term accounts for only about $10\%$ of the total input.
A meaningful solution, however, requires the input to capture the dominant dynamics.
Indeed, the LO term, which carries the dominant dynamics, already yields a reasonable solution when used alone, as shown in Fig.~\ref{fig:Convergence}(a).
By contrast, taking only the vacuum-condensate terms as input is even less satisfactory, since $f_{a_1}^2$ turns negative and no reasonable $f_{a_1}$ can be obtained.

Furthermore, the uniqueness of the inverse problem is also reflected in Fig.~\ref{fig:Convergence}.
Using the same integral equation, Eq.~\eqref{eq:final_equation}, but with different inputs, we obtain different results, as shown in Figs.~\ref{fig:Convergence}(a)--(d).

\subsection{Quark-Hadron Duality}\label{sec:quark-hadron-duality}

As discussed in the preceding section, the spectral function is nearly flat and agrees closely with its OPE counterpart over $5$~GeV$^2 \lesssim s\lesssim\Lambda$, as seen in Fig.~\ref{fig:final_result}.
This matching is a manifestation of quark-hadron duality: the equivalence between the quark-level description given by the OPE and the hadronic dynamics encoded in the reconstructed spectral function at large $s$.
In this subsection, we further discuss quark-hadron duality.

In the inverse problem approach, duality enters only as an input assumption for $s\geqslant\Lambda$, as in  Eqs.~(\ref{eq:original_inverse_problem}) and (\ref{eq:final_dispersion_relation}), where the spectral function in the integral over $[\Lambda,\infty)$ is identified with its OPE counterpart, $\rho(s)=\rho^{\rm OPE}(s)$.
For $t_{\rm min}\leqslant s\leqslant \Lambda$, by contrast, the spectral function is determined entirely by the inverse problem approach, without assuming duality.
Nevertheless, the resulting spectral function automatically agrees with the OPE result over $5$~GeV$^2 \lesssim s\lesssim\Lambda$ (with $\Lambda\simeq5\text{--}10$~GeV$^2$), which provides a check on the output that quark-hadron duality begins to hold at $s\approx 5~\mathrm{GeV}^2$.
The smooth matching between the duality obtained from the inverse problem below $\Lambda$ and the duality imposed as an input above $\Lambda$ demonstrates the self-consistency of the framework of the inverse problem approach.
Since this transition between the perturbative and non-perturbative regimes is continuous, $\Lambda$ should be interpreted as lying in the region where quark-hadron duality sets in, rather than as marking a unique boundary point.
Across this region, the solution is insensitive to variations of $\Lambda$, exhibiting a stability plateau, while quark-hadron duality holds on both sides of $\Lambda$.

In QCD sum rules, the hadronic spectral function is generally parameterized by a ground-state pole represented by a delta function and a continuum introduced through a step function~\cite{Shifman:1978bx,Shifman:1978by,Colangelo:2000dp,Khodjamirian:2020btr},
\begin{equation}
\rho(s)
=
f_H^2\,\delta(s-m_H^2)
+
\rho^{c}(s)\,\theta(s-s_{\rm th}^H)\,,
\end{equation}
which is similar to Eq.~(\ref{eq:spectral_decompose_N1}).
However, the continuum spectrum is not determined within the QCD sum rules approach.
Instead, quark-hadron duality is imposed just above the hadronic threshold $s_{\rm th}^H$.
Specifically, for an un-subtracted dispersion relation, the global quark-hadron duality ansatz reads
\begin{equation}\label{eq:global_duality}
    \int_{s_{\rm th}^H}^{\infty} ds \, \frac{\rho^c(s)}{s-q^2} \simeq \int_{s_0}^{\infty}ds\, \frac{\frac{1}{\pi} \operatorname{Im}\Pi^{\rm pert}(s)}{s-q^2}\,.
\end{equation}
Consequently, the hadronic spectral integral over $s\in[s_{\rm th}^H,\infty)$ is replaced by the perturbative integral over $s\in[s_0,\infty)$, where $\operatorname{Im}\Pi^{\rm pert}(s)$ is known.
Then, the only remaining unknowns in the sum rules are the ground-state residue $f_H^2$ and mass $m_H^2$, once $s_0$ is given.
The physical threshold $s_{\rm th}^H$ typically lies around the first excited state in QCD sum rules~\cite{Colangelo:2000dp}.
The region near $s_{\rm th}^H$ is thus entirely non-perturbative, where a perturbative description is not reliable.
Indeed, as seen in Fig.~\ref{fig:final_result}, the hadronic spectral function deviates substantially from the OPE spectrum around $a_1(1260)$. 
Since the perturbative description fails near $s_{\rm th}^H$, the lower limit of the perturbative integral must be shifted to $s_0\neq s_{\rm th}^H$ in order to match the hadronic and perturbative integrals and thereby implement global quark-hadron duality.
Consequently, $s_0$ becomes a process-dependent phenomenological parameter that cannot be determined from first principles, and the extracted predictions are often sensitive to its value.
In some physical applications, the uncertainty associated with $s_0$ constitutes the dominant source of error in QCD sum rules predictions.

We now compare the effective threshold $s_0$ in QCD sum rules with the separation scale $\Lambda$ in the inverse problem approach. 
Formally, both serve as the lower limits of the spectral integrals over which quark-hadron duality is imposed. 
In practice, however, $s_0$ typically lies in the low-energy non-perturbative region, whereas $\Lambda$ is chosen in a regime where perturbation theory is reliable. 
More specifically, in QCD sum rule analyses based on $\Pi_q(q^2)$, one finds $s_0 \simeq 0.7~\mathrm{GeV}^2$ when $f_\pi$ is extracted without an explicit contribution of excited states~\cite{Colangelo:2000dp}, whereas including the $a_1(1260)$ contribution raises the effective threshold to $s_0 \simeq (1.3\text{--}1.7)~\mathrm{GeV}^2$~\cite{Jiang:2015paa}. 
By contrast, the extraction of $f_{a_1}$ from $\Pi_g(q^2)$ employs $s_0 = 2.55 \pm 0.15~\mathrm{GeV}^2$~\cite{Yang:2007zt}. 
These examples show that the value of $s_0$ depends on the physical quantity being extracted, the choice of correlation function, and the number of states included in the spectral function.
Consequently, $s_0$ must generally be determined separately for each analysis, making it a phenomenological parameter to which the extracted results are highly sensitive. 
The underlying reason is that, in the low-energy region, QCD sum rules invoke quark-hadron duality and thereby avoid an explicit description of highly excited states and continuum contributions when extracting properties of the lowest-lying states.

The inverse problem approach enables the simultaneous reconstruction of the entire non-perturbative spectrum, including the ground state, excited states, and the continuum.
This ability to reconstruct the continuum is what fundamentally distinguishes the inverse problem approach from QCD sum rules.
Accordingly, quark-hadron duality is invoked only in the region where perturbation theory is reliable.
More specifically, we take $\Lambda\in[5,10]~\mathrm{GeV}^2$, which is substantially larger than the values of $s_0$ typically adopted in the analyses of QCD sum rules.
Moreover, the results exhibit negligible sensitivity to variations of $\Lambda$ within this range.
In contrast to QCD sum rules which rely on the global duality ansatz of Eq.~(\ref{eq:global_duality}), the inverse problem approach employs local quark-hadron duality for $s\gtrsim\Lambda$, where the hadronic and the OPE spectral functions are matched point by point.
Note that the duality in QCD sum rules concerns only the perturbative part without vacuum condensates, $\Pi^{\rm pert}(s)$, whereas in the inverse problem it concerns the OPE result, $\Pi^{\rm OPE}(s)$, which constitutes the full quark-level calculation. 
The inverse problem approach thereby avoids the ambiguities associated with the global duality ansatz and the effective threshold $s_0$, resulting in smaller and better-controlled uncertainties.

Our result that quark-hadron duality is applicable for $s\gtrsim 5~\mathrm{GeV}^2$ agrees with the conclusions of several previous studies.
Ref.~\cite{Narison:2024uud} argues that a physically meaningful duality threshold $s_0$ should lie in a region where resonance peaks are no longer prominent and perturbative QCD dominates.
In this spirit, the duality condition in $e^+e^-\to\mathrm{hadrons}$ requires $s_0\simeq(5.1\text{--}5.6)~\mathrm{GeV}^2$~\cite{Narison:2024uud}, a value supported by the experimental measurements of the $R$ ratio, which show an approximately flat behavior consistent with perturbative QCD from $(2.2324~\mathrm{GeV})^2\simeq 5.0$~GeV$^2$ up to the $J/\psi$ resonance region~\cite{BESIII:2021wib}.
This value of $s_0$ is in good agreement with the range of $\Lambda$ found in our analysis.
In addition, quark-hadron duality also plays a central role in the theoretical analyses of inclusive decays and lifetimes of heavy hadrons, as well as in neutral heavy-meson mixing~\cite{Shifman:2000jv,Bigi:2001ys}.
The duality is generally reliable for $B$ mesons but can break down for $D$ mesons~\cite{Jubb:2016mvq}, a pattern naturally understood from $m_b^2\simeq17~\mathrm{GeV}^2\gg\Lambda$ and $m_c^2\simeq1.7~\mathrm{GeV}^2\ll\Lambda$.

\begin{figure}
    \centering
    \includegraphics[width=0.4\linewidth]{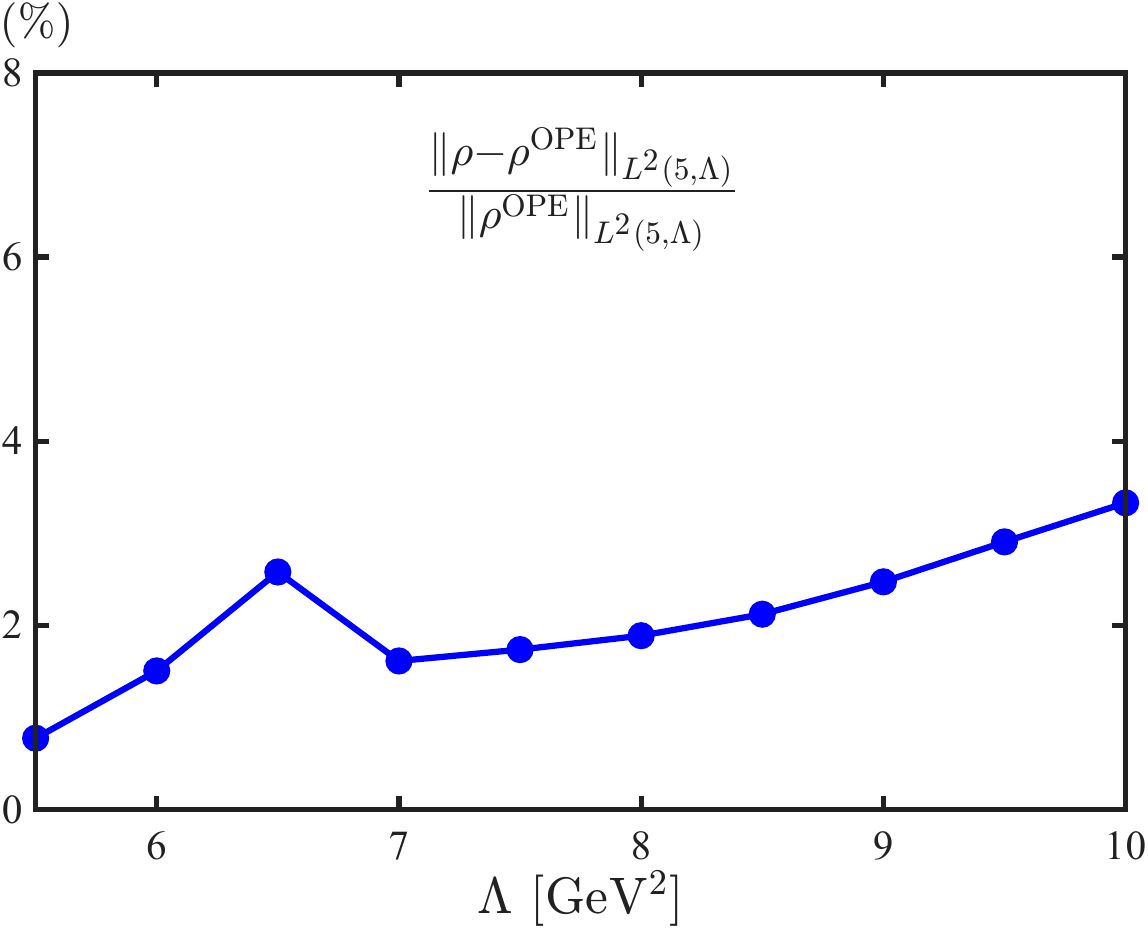}
    \caption{Quark-hadron duality violation with respect to $\Lambda$.}
    \label{fig:Quark-hadron_duality}
\end{figure}

To provide a more direct and quantitative measure of quark-hadron duality or its violation, we evaluate the relative deviation of the reconstructed spectrum $\rho(s)$ from its OPE counterpart $\rho^{\rm OPE}(s)$ for $s\geqslant 5~\mathrm{GeV}^2$.
Specifically, we quantify this deviation through the ratio of norms,
\begin{equation}
\frac{\left\|\rho-\rho^{\rm OPE}\right\|_{L^2(5,\Lambda)}}{\left\|\rho^{\rm OPE}\right\|_{L^2(5,\Lambda)}}\,,
\end{equation}
where the $L^2$ norm is evaluated over the interval $[5~\mathrm{GeV}^2,\Lambda]$, and $\Lambda$ is varied from $5.5$ to $10~\mathrm{GeV}^2$.
The result is shown in Fig.~\ref{fig:Quark-hadron_duality}.
It can be seen that the violation of quark-hadron duality is at the level of $(1\text{--}4)\%$ over this range of $s$.
This is consistent with the expectation that quark-hadron duality violations are at the few-percent level~\cite{Shifman:2000jv}.

As emphasized in the preceding section, a smooth agreement between the reconstructed spectral function and the OPE spectrum at $s\gtrsim5$~GeV$^2$ provides an important criterion for assessing the reliability of the solution.
The present analysis quantifies this criterion by requiring that, over a finite interval $s\in[5~\mathrm{GeV}^2,\Lambda]$, the two spectra approach each other smoothly, and the quark-hadron duality violation remains below approximately $5\%$.
We stress, however, that this level of duality violation is not imposed to determine any parameters but serves solely to assess the quality of the reconstructed solution.

\subsection{Borel Transformation}
To reduce sensitivity to the quark-hadron duality approximation and improve its accuracy, QCD sum rules commonly employ a Borel transformation~\cite{Shifman:1978bx,Shifman:1978by,Colangelo:2000dp,Khodjamirian:2020btr}.
After this transformation, the duality relation becomes
\begin{equation}
    \int_{s_{\rm th}^H}^{\infty} ds\,
    \rho^c(s)\,e^{-s/M^2}
    \simeq
    \frac{1}{\pi}\int_{s_0}^{\infty} ds\,
    \textbf{Im}\Pi^{\rm pert}(s)\,e^{-s/M^2}\,,
\end{equation}
where the Borel parameter $M^2$ must be determined within an appropriate window.
This window is determined by the simultaneous requirements of OPE convergence and pole dominance.
In practice, these requirements are implemented by prescribing two numerical criteria for the relative sizes of the relevant contributions.
On the one hand, the contribution from highly excited states and the continuum must remain sufficiently small relative to the lowest-lying state poles, ensuring pole dominance and reducing sensitivity to the quark–hadron duality approximation.
On the other hand, the condensate contribution is required to remain sufficiently small to ensure the convergence and reliability of the OPE.
However, the specific percentages assigned to these two criteria are not uniquely determined, and their choices inevitably introduce some arbitrariness and potential bias.
Moreover, whether the pole contribution itself constitutes a physically meaningful quantity remains a matter of debate~\cite{Zhang:2026roz}, since its value is highly sensitive to the unphysical Borel parameter $M^2$.
Recent studies have sought to reduce this arbitrariness within the QCD sum rules framework, for example through improved moment sum rules~\cite{Zhang:2026roz} and Bayesian approaches based on the maximum entropy method~\cite{Gubler:2010cf}.

\begin{figure}
                \centering
                {\includegraphics[width=0.4\textwidth ]{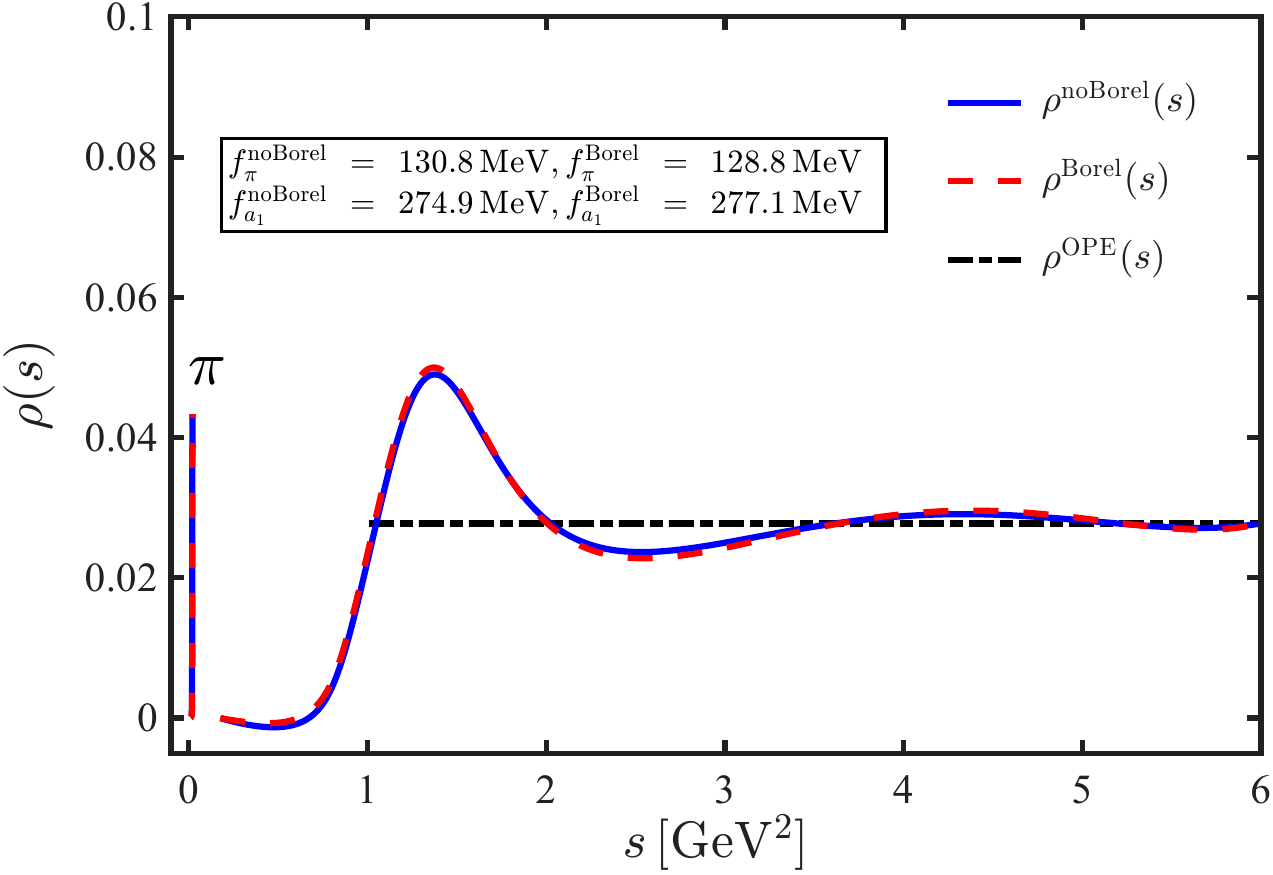}}
            
            \caption{Comparison between results with and without the Borel transformation.}
            \label{fig:Borel_transformation_vs_dispersion_relation}
\end{figure}

In this work, we also formulate and solve the inverse problem based on the Borel-transformed integral equation,
\begin{equation}
    \int_{t_{\min}}^{\Lambda} ds\,
    e^{-s/M^2}\rho(s)
    =
    \int_0^{\Lambda} ds\,
    e^{-s/M^2}\rho^{\mathrm{OPE}}(s)
    +
    \Pi^{\mathrm{cond}}(M^2)\,.
\end{equation}
The corresponding result is shown in Fig.~\ref{fig:Borel_transformation_vs_dispersion_relation}.
As can be seen, the decay constants of the pion and $a_1(1260)$ change by $2$~MeV or less, and the full spectral functions almost perfectly overlap, indicating that the reconstructed result is essentially unchanged regardless of whether the Borel transformation is applied.
The analysis proceeds analogously to that based on the dispersion-relation integral equation, Eq.~(\ref{eq:final_equation}), with the integral kernel replaced by $e^{-s/M^2}$.
Rather than searching for a stability plateau as a function of $M^2$, we employ a finite interval in $M^2$ as the input domain for the inverse-problem reconstruction.
Specifically, we take $M^2\in[5,100]~\mathrm{GeV}^2$ and find clear stability plateaus when the lower and upper bounds are varied independently.
The upper bound can be chosen arbitrarily large and, in principle, extended to infinity, whereas the lower bound is constrained by the requirement that the OPE remains convergent and reliable.
These two bounds play roles analogous to $q_{\rm low}^2$ and $q_{\rm up}^2$ in the dispersion-relation formulation, respectively.
In addition, a plateau in $\Lambda$ also appears in the Borel-transformed case.

In summary, the comparison between the results with and without the Borel transformation demonstrates the consistency of these two scenarios within the inverse problem approach.
Unlike in QCD sum rules, the Borel transformation is not necessary in the inverse problem approach. The excited states and the continuum are reconstructed explicitly, and quark-hadron duality emerges dynamically in the perturbative region. 
Consequently, there is no need to use the Borel transformation to suppress these contributions or to reduce the sensitivity to the quark-hadron duality approximation.
Moreover, we impose no prescribed numerical criteria on the relative sizes of the condensates and the pole contributions, thereby avoiding the arbitrariness and uncertainties inherent in such criteria.

\subsection{Decomposition of Spectral Function}\label{sec:condensates}

In this subsection, we discuss the contributions of each term in the OPE input to the ground state, excited states, and the continuum.

As can be seen from Eq.~(\ref{eq:OPE_calculation}), the input in this work is the same as that in QCD sum rules, but the procedure for extracting the non-perturbative quantities differs.
Within QCD sum rules, the phenomenological spectral function is truncated to the lowest-lying state, while the highly excited states and the continuum are replaced by the perturbative QCD spectral density through the quark–hadron duality approximation.
Accordingly, the condensate contributions enter through the OPE side, and their effects are reflected in the extracted parameters of the lowest-lying state.

By contrast, the inverse problem approach resolves the ground state, the excited states, and the continuum simultaneously.
The linearity of the inverse problem permits a decomposition of the reconstructed solution, as detailed in Appendix~\ref{sec:Minimization_functional}.
By solving the linear system separately for each input component, namely the perturbative term and each condensate contribution, we can quantify their individual contributions to the pion, the $a_1(1260)$ resonance, and the continuum. 
These three components are given by
\begin{equation}
    \int_{t_{\min}}^{\Lambda} ds\,\rho(s)
    =
    f_\pi^2
    +
     f_{a_1}^2\,\eta
    +
    \int_{s_{\rm th}}^{\Lambda} ds\,\rho^c(s)\,,
\end{equation}
with $\eta \equiv \int_{s_{\rm th}}^{\Lambda} ds\,\mathbb{BW}\left(s,M_{a_1(1260)}, \Gamma_{a_1(1260)} \right) =0.80$. 
The results are summarized in Table~\ref{tab:Decompose}. 
The three columns correspond to $f_\pi^2$, $f_{a_1}^2\,\eta$, and $\int_{s_{\rm th}}^\Lambda ds\,\rho^c(s)$.
Each row corresponds to an individual term in the OPE input of Eq.~(\ref{eq:OPE_calculation}).
For each column, the first entry is the reconstructed value in units of $\mathrm{GeV}^2$, and the value in parentheses is its percentage share of the corresponding quantity.
\begin{table}
    \centering
    \caption{Contributions from each term in the OPE input (rows) to the reconstructed spectral weights of the pion, the $a_1(1260)$ resonance, and the continuum (columns). For each column, the first entry denotes the reconstructed value in $\mathrm{GeV}^2$, and the parenthetical value denotes its percentage share of the corresponding quantity.}
    \label{tab:Decompose}
    \begin{tabular}{@{}l|r@{\ }r|r@{\ }r|r@{\ }r@{}}
    \hline
    \hline
     & \multicolumn{2}{c|}{\(f_{\pi}^2\)}
     & \multicolumn{2}{c|}{\(f_{a_1(1260)}^2\eta \) }
     & \multicolumn{2}{c}{\(\int_{s_{th}}^\Lambda ds\,\rho_i^c(s)\)} \\
    \hline
      pure LO
    &   \(0.011\) &  \((66.6\%)\)
    &   \(0.065\) &  \((108.6\%)\)
    &   \(0.075\) &  \((84.3\%)\) \\
     pure NLO
    &   \(0.001\) &  \((6.3\%)\)
    &   \(0.006\) &  \((10.3\%)\)
    &   \(0.007\) &  \((8.0\%)\) \\
    \(m_q\left<\bar{q} q\right>\)
    &  \(-0.0002\) &  \((-1.3\%)\)
    &  \(0.0005\) &  \((0.8\%)\)
    &  \(-0.0003\) &  \((-0.3\%)\) \\
    \(\left<\alpha_s G^2 \right>\)
    &  \(0.003\) &  \((18.2\%)\)
    &  \(-0.007\) &  \((-11.5\%)\)
    &  \(0.004\) &  \((4.2\%)\) \\
    \(\left<\bar{q}q \bar{q}q \right>\)
    &  \(0.002\) &  \((9.9\%)\)
    &  \(-0.005\) &  \((-8.3\%)\)
    &  \(0.003\) &  \((3.7\%)\) \\
    \hline
    Sum 
    &  \(0.017\) &   \((100.0\%)\)
    &  \(0.060\) &   \((100.0\%)\)
    &  \(0.089\) &   \((100.0\%)\) \\
    \hline
    \hline
    \end{tabular}
\end{table}
For this decomposition, we fix the regularization parameter $\alpha$ at the value determined from Fig.~\ref{fig:Convergence}(e).
We can see that, for each of the three components of the reconstructed solution, the LO perturbative term gives the dominant contribution, followed by the NLO contribution, the gluon condensate, and the four-quark condensates.
The contributions from the quark condensate and the mixed quark-gluon condensate are small since they are suppressed by the light quark masses.
This hierarchy also demonstrates the convergence of the OPE.
Moreover, Table~\ref{tab:Decompose} clearly shows that the condensate terms contribute not only to the ground-state residue but also to the excited state and the continuum.
This highlights another important difference between the inverse problem approach and QCD sum rules.


Furthermore, Table~\ref{tab:Decompose} shows that, for each condensate, the contributions to $f_\pi^2$, $f_{a_1}^2\,\eta$, and $\int_{s_{\rm th}}^\Lambda ds\,\rho^c(s)$ sum to zero within numerical accuracy. 
This cancellation reflects the superconvergence of the individual condensate contributions.
To see this, recall that each condensate term
$\Pi_i^{\rm cond}(q^2)$ in Eq.~\eqref{eq:OPE_calculation} behaves as
$1/(q^2)^2$ or $1/(q^2)^3$ at large spacelike momentum, with no $1/q^2$
contribution. Consequently, it satisfies the superconvergence condition
\begin{equation}
    \lim_{q^2 \to -\infty} q^2 \Pi_i^{\rm cond}(q^2) = 0.
    \label{eq:superconvergence}
\end{equation}
Because of this falloff, no subtraction polynomial is required, and the dispersion relation takes the unsubtracted form
\begin{equation}
    \Pi_i^{\rm cond}(q^2) = \int_0^\infty ds\,\frac{\rho_i(s)}{s-q^2},
    \label{eq:unsubtracted_i}
\end{equation}
where $\rho_i(s)$ is the spectral function associated with the $i$-th condensate term. 
We emphasize that Eq.~\eqref{eq:unsubtracted_i} holds in the sense of distributions.
Multiplying both sides of Eq.~\eqref{eq:unsubtracted_i} by $q^2$ and taking the limit $q^2\to-\infty$, the left-hand side vanishes by Eq.~\eqref{eq:superconvergence}. On the right-hand side, the kernel $ K_{q^2}(s)=\frac{q^2}{s-q^2}$ converges to $-1$ as $q^2\to-\infty$ for every fixed $s\geqslant 0$. 
Moreover, for $q^2<0$ and $s\geqslant 0$, one has $|K_{q^2}(s)|=\frac{|q^2|}{s+|q^2|}\leqslant 1$,
i.e. the kernel is uniformly bounded in $s\geqslant0$ for all $q^2<0$. 
Then the limit and the integral can be interchanged, so that $\lim_{q^2\to-\infty}\int_0^\infty ds\,\rho_i(s)K_{q^2}(s)
    = -\int_0^\infty ds\,\rho_i(s)$.
Combining this with the vanishing of the left-hand side, we obtain 
\begin{equation}
    0 = \int_0^\infty ds\,\rho_i(s).
    \label{eq:zeromoment}
\end{equation}

Decomposing the spectral function into the pion pole, the $a_1$ contribution, the low-energy continuum, and the high-energy OPE tail, and substituting into Eq.~(\ref{eq:zeromoment}), one obtains
\begin{equation}
\begin{split}
f_{\pi,i}^{\,2}
+
f_{a_1,i}^{\,2}\,\eta
+
\int_{s_{\rm th}}^\Lambda \rho_i^c(s)\,ds
=-
\int_\Lambda^\infty
\rho_i^{\mathrm{OPE}}(s)\,ds .
\end{split}
\label{eq:condensate-superconvergence}
\end{equation}
For the quark, gluon, and four-quark condensates, the OPE spectral density $\rho_i^{\rm OPE}$ vanishes for $s\in [\Lambda,\infty)$. 
Consequently, the three terms on the left-hand side of Eq.~\eqref{eq:condensate-superconvergence} sum to zero, in agreement with the numerical result of Table~\ref{tab:Decompose}.

\section{Summary and Outlook}\label{sec:summary}

This work applies the inverse-problem framework of Ref.~[1] for the first time to a realistic hadronic observable. 
The axial-vector correlation function and its dispersion relation are used to simultaneously determine the pion decay constant, the $a_1(1260)$ decay constant, and the continuum spectral density. 
The results are satisfactory on both theoretical and phenomenological grounds. 
On the theoretical side, the method is self-consistent. 
The results exhibit stability plateaus with respect to all method-intrinsic parameters, namely the regularization parameter $\alpha$, the separation scale $\Lambda$, the subtraction point $\sigma_0$, and the upper and lower bounds of the $q^2$ interval. 
Moreover, the reconstructed solution agrees closely with the OPE spectrum at the high-energy end of the solution interval without any additional constraints.
On the phenomenological side, the low-energy results agree with experimental data.
The extracted $f_\pi = 130.8 \pm 8.1~\mathrm{MeV}$ is consistent with the measured $130.57 \pm 0.14~\mathrm{MeV}$ within uncertainties. 
The reconstructed spectral density agrees well with the ALEPH $\tau$-decay data over almost the entire range, covering both the $a_1(1260)$ peak and the continuum. 

The inverse problem approach presented in this work offers four main advantages. 
First, it is mathematically rigorous. 
The inversion is ill-posed, admitting a unique but unstable solution, and Tikhonov regularization renders it stable while guaranteeing that the regularized solution converges to the exact one as the input error $\delta \to 0$. 
The precision of the extraction can therefore be improved systematically with better inputs. 
Second, it allows for a detailed error budget. 
Uncertainties are separated into input-induced and method-intrinsic contributions. 
In particular, the approximation error due to the finite regularization parameter $\alpha$ can be estimated through $\|R_\alpha K f - f\|$. 
In the present analysis, the method-intrinsic uncertainties are well below the input-induced ones, which shows that the method itself introduces no significant bias. 
Third, it solves for the ground state, the excited states, and the continuum simultaneously within a single framework. 
The ability to reconstruct the continuum spectrum is a distinctive advantage of the inverse problem approach.
Fourth, it naturally realizes quark-hadron duality at high energy scales.
This avoids the low-energy duality assumption and the Borel transformation, thereby eliminating the uncertainties associated with them.

Several further physics analyses are performed. 
We first verify the convergence of the solutions with improving input precision.
As the input is upgraded from pure LO to LO$+$condensates and then to the full LO$+$NLO$+$condensates, the extracted $f_\pi$ successively approaches the experimental value, and the spectral density agrees increasingly well with both the data and the OPE result. 
We then study quark-hadron duality. 
It is assumed as an input only for $s \geqslant \Lambda$, while the region below $\Lambda$ is determined entirely by the inversion.
Nevertheless, the reconstructed density is found to match the OPE curve over $5~\mathrm{GeV}^2 \lesssim s \lesssim 10~\mathrm{GeV}^2$, with the duality violation estimated to be only $(1\text{--}4)\%$ in this range. 
We further find that the results are essentially unchanged with or without the Borel transformation. 
The decay constants shift by less than $1~\mathrm{MeV}$ and the spectral functions almost coincide. 
Therefore, the method does not rely on the Borel transformation and avoids the arbitrariness associated with the choice of the Borel parameter. 
Finally, exploiting the linearity of the problem, the spectral density is decomposed into the individual OPE-term contributions to the pion, the $a_1(1260)$, and the continuum. 
For every condensate term, the three partial contributions cancel to zero within numerical accuracy, explicitly demonstrating their superconvergence.

Broad physical applications of the inverse problem approach can be expected in the future. 
For experimentally well-established hadronic states such as the pion and $a_1(1260)$, one can study their inner structures, including light-cone distribution amplitudes (LCDAs), parton distribution functions (PDFs), generalized parton distributions (GPDs), and so on. 
These quantities can be obtained by replacing the interpolating currents and correlation functions used in this work with those for the corresponding moments.
The inner structure functions can then be extracted from these moments.
There would be no essential difference between the present calculations and those for the moments.

A major advantage of the inverse problem approach is that it allows one to reconstruct the continuum contribution and thereby obtain the full spectral density, a capability that enables a wide range of applications.
The hadronic vacuum polarization contribution to the muon anomalous magnetic moment requires the spectral function over the entire energy range~\cite{Aoyama:2020ynm}.
The Weinberg sum rules relate integrals of the difference between the vector and axial spectral functions, $\int_0^\infty ds\, \left[ \rho_V(s)-\rho_A(s) \right] = f_{\pi}^2$ and $\int_0^\infty ds \,s \left[ \rho_V(s)-\rho_A(s) \right]= 0$, with the continuum, resonance, and pion-pole contributions required to jointly satisfy these constraints~\cite{Weinberg:1967kj}. 
In the extraction of precision QCD parameters, the continuum spectral function plays an essential role, as it enters the weighted integrals that define the spectral moments used to determine $\alpha_s$ and the low-energy constants of chiral perturbation theory~\cite{ALEPH:2005qgp,Gonzalez-Alonso:2008sva,Pich:2013lsa}.
Spectral functions at finite temperature and density determine transport coefficients, such as shear viscosity and electrical conductivity~\cite{Meyer:2011gj}.

\section*{Acknowledgments}
We are very grateful to Dong-Hao Li, Hsiang-nan Li and Ji Xu for many fruitful discussions throughout this project.
This work is supported by the Scientific Research Innovation Capability Support Project for Young Faculty under Grant No. ZYGXQNJSKYCXNLZCXM-P2,  by the Fundamental Research Funds for the Central Universities under No. lzujbky-2023-stlt01, lzujbky-2024-oy02 and lzujbky-2025-eyt01, and the National Natural Science Foundation of China under Grant No.12571455, and No.12335003.

\appendix
\section{Proof of Uniqueness and Instability of Inverse Problem with a Pole}\label{sec:proof_illposedness}

We consider the inverse problem for the dispersion relation with a pole, as formulated in Eq.~(\ref{eq:final_equation})
\begin{equation}{\label{eq:our-case-dispersion-relation}}
\frac{f_\pi^2}
{(m_\pi^2-\sigma_0)(m_\pi^2-q^2)}
+
\int_{s_{\rm th}}^\Lambda ds\,
\frac{\rho^c(s)}
{(s-\sigma_0)(s-q^2)}
=g(q^2).
\end{equation}
Here $q^2 \in [q^2_{\rm low}, q^2_{\rm up}]$ with $q^2_{\rm low}< q^2_{\rm up}\ll0$, $0 < m_\pi^2 < s_{\rm th} < \Lambda$, $\rho^c\in F=L^2(s_{\rm th},\Lambda)$ and $g\in G=L^2(q_{\rm low}^2, q_{\rm up}^2)$. 
In order to satisfy the perturbation of OPE, we require $\sigma_0< q^2_{\rm low}$ and $|q^2_{\rm up}|>\Lambda$.
In particular, the denominators $m_\pi^2-\sigma_0$, $s-\sigma_0$, $m_\pi^2-q^2$, and $s-q^2$ are nonzero and positive in the relevant domains.

\begin{proof}[{\bf Proof of uniqueness}]
To prove uniqueness, suppose there are two solutions.
Their difference satisfies the homogeneous equation.
Thus uniqueness is established if we can show that the homogeneous equation admits only the trivial solution, i.e., $g(q^2)=0$ implies $\rho^c(s)=0$ and $f_\pi^2=0$.
It suffices to consider the homogeneous equation
\begin{equation}
\frac{f_\pi^2}
{(m_\pi^2-\sigma_0)(m_\pi^2-q^2)}
+
\int_{s_{\rm th}}^\Lambda ds\,
\frac{\rho^c(s)}
{(s-\sigma_0)(s-q^2)}
=0 .
\end{equation}
Let $u=q^2-m_\pi^2$, $t=s-m_\pi^2$, and $a=s_{\rm th}-m_\pi^2$, $b=\Lambda-m_\pi^2$, with $0<a\leqslant t \leqslant b$, $u<0$.
Define $H(t)$ by
\begin{equation}
H(t)=\frac{m_\pi^2-\sigma_0}{t+m_\pi^2-\sigma_0}\rho^c(t+m_\pi^2).
\end{equation}
which belongs to $L^2$ space. Since $t+m_\pi^2-\sigma_0=s-\sigma_0>0$ and $m_\pi^2-\sigma_0>0$, the factor $(m_\pi^2-\sigma_0)/(t+m_\pi^2-\sigma_0)>0$, so $H(t)=0$ if and only if $\rho^c(s)=0$.
Multiplying $m_\pi^2-\sigma_0$, the homogeneous equation becomes
\begin{equation}\label{eq:simplified-homogeneous-equation}
\frac{f_\pi^2}
{u}
+
\int_a^b dt\,
\frac{H(t)}{u-t}
=0 .
\end{equation}
Since $q_{\rm up}^2\leqslant -\Lambda$, we have $u=q^2-m_\pi^2\leqslant q_{\rm up}^2-m_\pi^2 \leqslant -\Lambda-m_\pi^2 < -(\Lambda-m_\pi^2)= -b$, so $|u|>b$. Since $t\leqslant b$, it follows that $|t/u|<1$.
Then, we have the expansion
\begin{equation}
\frac{1}{u-t}
=
\frac{1}{u}\sum_{k=0}^{\infty}\left(\frac{t}{u}\right)^k .
\end{equation}
Therefore, $\frac{f_\pi^2}{u} + \sum_{k=0}^{\infty} \frac{1}{u^{k+1}} \int_a^b dt\,t^k H(t) =0 $.
Multiplying by $u$, we obtain
\begin{equation}\label{eq:expanded-u}
f_\pi^2
+
\sum_{k=0}^{\infty}
\frac{1}{u^{k}}
\int_a^b dt\,t^k H(t)
=0 .
\end{equation}
This identity holds for all $u$ in a non-degenerate interval of the negative real axis. Define the moments $m_k=\int_a^b dt\,t^k H(t)$, $k=0,1,2,\dots$.
Then \eqref{eq:expanded-u} becomes
\begin{equation}\label{eq:moment-series}
f_\pi^2+\sum_{k=0}^{\infty}\frac{m_k}{u^k}=0.
\end{equation}
Since this holds on a non-degenerate interval of $u$, all coefficients of the Laurent expansion in $u$ vanish. Hence
\begin{equation}\label{eq:moment-conditions}
f_\pi^2+m_0=0,
\qquad\text{and}\qquad
m_k=0,\quad k=1,2,3,\dots.
\end{equation}
This can also be obtained by successively multiplying $u^k \,(k=0,1,2,\cdots)$ and letting $u\to -\infty$.
At this stage, we know that $H$ is orthogonal to all polynomials $t^k$ with $k\geqslant 1$, but the $k=0$ moment is not yet known to vanish because $f_\pi^2$ may be nonzero. To proceed, we use analytic continuation.

Consider the function
\begin{equation}\label{eq:Phi}
\Phi(u)=
\frac{f_\pi^2}{u}
+
\int_a^b dt\,\frac{H(t)}{u-t}.
\end{equation}
Let $D=\mathbb C\setminus\bigl([a,b]\cup\{0\}\bigr)$ i.e., the complex plane with the real interval $[a,b]$ and the point $0$ removed.
The function $\Phi$ is analytic on $D$. 
Indeed, the first term is analytic away from $u=0$, and $0\notin[a,b]$ because $a>0$, while the integral term is analytic off the cut $[a,b]$ by standard properties of the Cauchy transform.

By Eq.~\eqref{eq:simplified-homogeneous-equation}, $\Phi(u)=0$ for all
$u\in[q_{\rm low}^2-m_\pi^2,\ q_{\rm up}^2-m_\pi^2]$,
which is a non-degenerate interval contained in $D$. 
Since $\Phi$ is analytic on $D$, the identity theorem for analytic functions implies that
\begin{equation}\label{eq:Phi-zero}
\Phi(u)=0
\qquad
\text{for all }u\in D.
\end{equation}

Now compute the discontinuity of $\Phi$ across the cut $[a,b]$. 
For almost every every $t_0\in(a,b)$, the Sokhotski--Plemelj formula for $H(t) \in L^2(a,b)$ gives $\frac{1}{t_0\pm i0-t}={\rm PV}\frac{1}{t_0-t}\mp i\pi\delta(t_0-t)$. 
Therefore,
\begin{equation}\label{eq:jump}
\Phi(t_0+i0)-\Phi(t_0-i0)
=
-2\pi i\,H(t_0),
\qquad t_0\in(a,b).
\end{equation}
Since $\Phi\equiv0$ on $D$, its boundary values from above and below coincide, $\Phi(t_0+i0)=\Phi(t_0-i0)=0$.
So the left-hand side of Eq.~\eqref{eq:jump} vanishes. Hence
\[
H(t_0)=0
\qquad\text{a.e. on }(a,b).
\]
Consequently,
\begin{equation}\label{eq:rho-zero}
\rho^c(s)
=
\frac{s-\sigma_0}{m_\pi^2-\sigma_0}
H(s-m_\pi^2)
=0
\qquad\text{a.e. on }[s_{\rm th},\Lambda].
\end{equation}
Substituting $H=0$ back into \eqref{eq:simplified-homogeneous-equation} gives
$f_\pi^2/u=0$, for all $u\in[q_{\rm low}^2-m_\pi^2,\ q_{\rm up}^2-m_\pi^2]$,
hence $f_\pi^2=0$. 
Therefore, the homogeneous equation admits only the trivial solution, and the solution of \eqref{eq:our-case-dispersion-relation} is unique.

\end{proof}

\begin{proof}[{\bf Proof of instability}]
We construct a counterexample. Let $L=\Lambda-s_{\rm th}$.
For a vector of exact solution $(f_\pi^2, \rho^c)$, define a noisy solution
\begin{equation}\label{eq:perturbed-solution}
f_{\pi,n}^{2}=f_\pi^2,\qquad
\rho^{c}_n(s)=\rho^c(s)+\sqrt n
\cos\left(\frac{n\pi(s-s_{\rm th})}{L}\right),
\end{equation}
where $n$ is an integer. 
Then
\begin{equation}\label{eq:perturbation-norm}
\begin{aligned}
\|(f_{\pi,n}^{2},\rho^{c}_n)-(f_\pi^2,\rho^c)\|_{L^2}^2
&=
\int_{s_{\rm th}}^\Lambda ds\,n\cos^2\left(\frac{n\pi(s-s_{\rm th})}{L}\right) =
\frac{nL}{2}
\xrightarrow[n\to\infty]{}\infty .
\end{aligned}
\end{equation}
Thus the perturbation diverges in norm.

Let $g_n$ be the data corresponding to $(f_{\pi,n}^{2},\rho^{c}_n)$. Since the pole part is unchanged,
\begin{equation}\label{eq:gn-g}
g_n(q^2)-g(q^2)
=
\int_{s_{\rm th}}^\Lambda ds\,
\frac{\sqrt n\cos\left(\frac{n\pi(s-s_{\rm th})}{L}\right)}
{(s-\sigma_0)(s-q^2)} .
\end{equation}
Because of $\sin 0=\sin(n\pi)=0$, integration by parts gives
\begin{equation}\label{eq:integration-by-parts}
\begin{aligned}
g_n(q^2)-g(q^2)
&=
\frac{L}{\sqrt{n}\,\pi\,(q^2 - \sigma_0)} \int_{s_{\rm th}}^{\Lambda} \sin\!\left(\frac{n\pi(s-s_{\rm th})}{L}\right) \left[ \frac{1}{(s-q^2)^2} - \frac{1}{(s-\sigma_0)^2} \right] ds
\\
&\leqslant \frac{L}{\sqrt{n}\,\pi\,(q^2 - \sigma_0)} \int_{s_{\rm th}}^{\Lambda}  \left[ \frac{1}{(s-q^2)^2} - \frac{1}{(s-\sigma_0)^2} \right] ds
\\
&=\frac{L^2}{\sqrt{n}\,\pi\,(q^2 - \sigma_0)} \left[ \frac{1}{(s_{\rm th} - q^2)(\Lambda - q^2)} - \frac{1}{(\Lambda - \sigma_0)(s_{\rm th} - \sigma_0)} \right]
\\
&\leqslant \frac{L^2}{\sqrt{n}\,\pi\,(q_{\rm low}^2 - \sigma_0)} \left[ \frac{1}{(s_{\rm th} - q_{\rm up}^2)(\Lambda - q_{\rm up}^2)} - \frac{1}{(\Lambda - \sigma_0)(s_{\rm th} - \sigma_0)} \right]
\\
&=\frac{L^2 C}{\sqrt{n}\,\pi\,}
\end{aligned}
\end{equation}
where $\sin(n\pi(s-s_{\rm th})/L)\leqslant 1$ and $\sigma_0< q_{\rm low}^2\leqslant q^2 \leqslant q_{\rm up}^2<0<s_{\rm th}<\Lambda$ are used, and set $C=\frac{1}{q_{\rm low}^2 - \sigma_0} \left[ \frac{1}{(s_{\rm th} - q_{\rm up}^2)(\Lambda - q_{\rm up}^2)} - \frac{1}{(\Lambda - \sigma_0)(s_{\rm th} - \sigma_0)} \right]>0$.
It follows that
\begin{equation}\label{eq:data-convergence}
\|g_n-g\|_{L^2}^2
\le
\frac{L^4C^2}{\pi^2}
(q_{\rm up}^2-q_{\rm low}^2)\,
\frac{1}{ n}
\xrightarrow[n\to\infty]{}0.
\end{equation}
Thus $g_n\to g$, while the corresponding solutions diverge in norm. Hence the inverse problem is unstable.
\end{proof}

\section{Minimization Procedure for the Tikhonov Functional}\label{sec:Minimization_functional}

In this appendix, we detail the minimization procedure for the Tikhonov functional, which reduces to solving a linear system. 
Unlike the case of a pure continuum spectral function considered in~\cite{Xiong:2022}, the present work includes pole and resonance contributions. 
Specifically, three cases are considered: a single pole, one pole associated with one resonance, and one pole with two resonances. 
Since the solution procedures for these cases are largely analogous, a detailed derivation is presented only for the case of one pole with one resonance. 
The other two cases are obtained by simply omitting or appending resonances as appropriate.

The full spectral function is decomposed into three parts: a pion pole described by a delta function, a finite-width $a_1(1260)$ resonance modeled by the Breit--Wigner propagator, and a continuum spectral function,
\begin{equation}
\rho(s) = f_\pi^2\,\delta(s-m_\pi^2) + f_{a_1}^2\,\mathbb{BW}_{a_1}(s) + \rho^c(s).
\end{equation}

The integral equation we are solving is
\begin{equation}
\frac{f_\pi^2}
{(m_\pi^2-\sigma_0)(m_\pi^2-q^2)}
+
f_{a_1}^2
\int_{s_{\rm th}}^\Lambda ds\,
\frac{\mathbb{BW}_{a_1}(s)}
{(s-\sigma_0)(s-q^2)}
+
\int_{s_{\rm th}}^\Lambda ds\,
\frac{\rho^c(s)}
{(s-\sigma_0)(s-q^2)}
=g(q^2),
\end{equation}
with $s_{\rm th}=(3m_\pi)^2$.
Note that in the second term, the integration over the Breit--Wigner propagator is a known result that depends on $q^2$, leaving only $f_{a_1}$ as the unknown parameter. 
This equation contains three unknown quantities: the pion residue $f_\pi^2$, the $a_1(1260)$ residue $f_{a_1}^2$, and the continuum spectrum $\rho^c(s)$.
We construct a dimensionless solution vector, $f=(f_\pi^2/m_\pi^2,\, f_{a_1}^2/m_{a_1}^2,\, \rho^c)$, which facilitates the subsequent solution. 

The inverse problem amounts to determining $f$ from the input $g(q^2)$. 
Because the inverse problem is ill-posed, Tikhonov regularization is adopted.
The Tikhonov functional for this problem is defined as 
\begin{equation}\label{eq:Tikhonov-appendix}
J_\alpha(f)=\left\|Kf -g^{\delta} \right\|^2_G+ \alpha \left\|\rho^c \right\|^2_F,
\end{equation}
where $\alpha$ is the regularization parameter, and $F=H^1(s_{\rm th},\Lambda)$ and $G=L^2(q_{\rm low}^2,q_{\rm up}^2)$ are the relevant function spaces.
The regularized solution is then obtained by minimizing the Tikhonov functional, $f_\alpha^\delta=\arg\min_f J_\alpha(f)$. 
Note that the penalty term includes only $\rho^c$, but not $f_\pi^2$ or $f_{a_1}^2$, since this term serves to regularize the ill-posed behavior of $\rho^c$. 
If $f_{a_1}^2$ were included, then, because $\mathbb{BW}_{a_1}(s)$ and $\rho^c(s)$ have the same integration range, the prior information on $a_1(1260)$ would be absorbed into $\rho^c(s)$ and could not be resolved separately. 

The numerical evaluation of the above integral equation usually requires discretization, which can follow the procedure in Section 4.4 of \cite{Xiong:2022}, but with additional pole and resonance terms.
The continuum function is expressed as
\begin{equation}
\rho^{c}_n(s)
=
\sum_{i=0}^{n}c_i\,\varphi_i(s),
\end{equation}
where $n$ is the number of grid points, and $\varphi_i(s)$ are piecewise basis functions in the interval $[s_{\rm th},\Lambda]$. 
We use the finite element method with hat functions, defined as
\begin{equation}\label{eq:finite-element}
\begin{aligned}
\varphi_i(x) &= 
\begin{cases} 
\dfrac{x - x_{i-1}}{h}, & x \in [x_{i-1}, x_i], \\[1mm]
-\dfrac{x - x_{i+1}}{h}, & x \in [x_i, x_{i+1}], \\[1mm]
0, & \text{otherwise},
\end{cases}
\\[2mm]
\varphi_0(x) &= 
\begin{cases} 
-\dfrac{x - x_1}{h}, & x \in [x_0, x_1], \\[1mm]
0, & \text{otherwise},
\end{cases}
\\[2mm]
\varphi_n(x) &= 
\begin{cases} 
\dfrac{x - x_{n-1}}{h}, & x \in [x_{n-1}, x_n], \\[1mm]
0, & \text{otherwise},
\end{cases}
\end{aligned}
\end{equation}
with $x_i=s_{\rm th}+ih$, $i=0,1,2,\ldots,n$, and $h=(\Lambda-s_{\rm th})/n$.
In practice, we take $h=0.01$ and set $\Lambda=s_{\rm th} + nh$. 
Two additional generalized basis functions are introduced to account for the pole and the resonance, 
\begin{equation}
\varphi_{-2}(s)
=
m_\pi^2\,\delta(s-m_\pi^2),
\qquad
\varphi_{-1}(s)
=
m_{a_1}^2\,\mathbb{BW}_{a_1}(s),
\end{equation}
with the corresponding coefficients 
\begin{equation}
c_{-2}
=
\frac{f_\pi^2}{m_\pi^2},
\qquad
c_{-1}
=
\frac{f_{a_1}^2}{m_{a_1}^2}.
\end{equation}
The complete spectral function is then expressed as
\begin{equation}
f_n(s)
=
\sum_{k=-2}^{n}c_k\,\varphi_k(s).
\end{equation}
Strictly speaking, the delta function is not an element of $L^2$, so $\varphi_{-2}$ should be regarded as a formal basis function introduced for physical convenience. 
Since it does not enter the penalty term, no integral involving $\delta^2$ arises. 

For convenience, we use the index $k, l=-2,-1,0,1,\ldots,n$ for the full spectral function, and $i,j=0,1,\ldots,n$ for its continuum part.

The integral operator acting on the basis functions then gives
\begin{equation}
\begin{aligned}
K\varphi_{-2}
&=
\frac{m_\pi^2}{(m_\pi^2-\sigma_0)(m_\pi^2-q^2)}\,,
\\
K\varphi_{-1}
&=
\int_{s_{\rm th}}^\Lambda ds\,
\frac{m_{a_1}^2\,\mathbb{BW}_{a_1}(s)}
{(s-\sigma_0)(s-q^2)}\,,
\\
K\varphi_i
&=
\int_{s_{\rm th}}^\Lambda ds\,
\frac{\varphi_i(s)}
{(s-\sigma_0)(s-q^2)},
\qquad i=0,\ldots,n.
\end{aligned}
\end{equation}
Note that all the integrals above are functions of $q^2$. 
Thus, the inverse problem can be compactly expressed as 
\begin{equation}
\sum_{k=-2}^{n} c_k\,\bigl(K\varphi_k\bigr)(q^2) = g(q^2).
\end{equation}
Thus the inverse problem is linear with respect to the coefficients $\{c_k\}$, which correspond to the pion pole, the finite-width $a_1(1260)$ resonance, and the continuum spectrum.

The Tikhonov functional in Eq.~(\ref{eq:Tikhonov-appendix}) is rewritten as
\begin{equation}
\begin{aligned}
J_\alpha(f_{\alpha,n})
&=
\left\| \sum_{k=-2}^n c_k\,K\varphi_k - g \right\|_G^2 + \alpha \left\| \sum_{i=0}^n c_i\,\varphi_i \right\|_F^2
\\
&=
\sum_{k,l=-2}^{n}
c_k c_l
( K\varphi_k, K\varphi_l)_G
-
2\sum_{k=-2}^{n}
c_k
( K\varphi_k, g)_G
+
( g, g)_G
+
\alpha
\sum_{i,j=0}^{n}
c_i c_j
(\varphi_i,\varphi_j)_F.
\end{aligned}
\end{equation}
Note that the last term does not include $c_{-2}$ and $c_{-1}$.  
The inner products are explicitly written as 
\begin{equation}
\begin{aligned}
(\varphi_i,\varphi_j)_F 
&= \int_{s_{\rm th}}^\Lambda \varphi_i(s) \,\varphi_j(s)\, ds, 
\\
( K\varphi_k, g)_G
&= \int_{q_{\rm low}^2}^{q_{\rm up}^2} \bigl(K\varphi_k\bigr)(q^2)\, g(q^2) \, dq^2,
\\
( K\varphi_k, K\varphi_l)_G
&=  \int_{q_{\rm low}^2}^{q_{\rm up}^2} \bigl(K\varphi_k\bigr)(q^2)\, \bigl(K\varphi_l\bigr)(q^2) \, dq^2.
\end{aligned}
\end{equation}

The coefficients $\{c_k\}$ are then determined by minimizing the Tikhonov functional, which is equivalent to requiring $\partial J_\alpha/\partial c_k = 0 $, $k=-2,-1,0,\ldots,n$.
The derivatives give
\begin{equation}
\sum_{l=-2}^{n}
\left[
( K\varphi_k,K\varphi_l)_G
+
\alpha \,B_{kl}
\right]c_l
=
(K\varphi_k,g)_G,
\end{equation}
where
\begin{equation}
B_{kl}
=
\begin{cases}
(\varphi_k,\varphi_l)_F\,, & k,l\geqslant 0,\\[1mm]
0, & k<0\ \text{or}\ l<0.
\end{cases}
\end{equation}

We define
\begin{equation}
A_{kl}
=
(K\varphi_k,K\varphi_l)_G\,,
\qquad
D_k
=
(K\varphi_k,g)_G\,,
\end{equation}
and collect the coefficients into
\[
\boldsymbol C
=
\begin{pmatrix}
c_{-2}\\
c_{-1}\\
c_0\\
\vdots\\
c_n
\end{pmatrix}
=
\begin{pmatrix}
f_\pi^2/m_\pi^2\\
f_{a_1}^2/m_{a_1}^2\\
c_0\\
\vdots\\
c_n
\end{pmatrix}.
\]
The matrix $\boldsymbol A$ is therefore
\[
\boldsymbol A
=
\begin{pmatrix}
A_{-2,-2} & A_{-2,-1} & A_{-2,0} & \cdots\\
A_{-1,-2} & A_{-1,-1} & A_{-1,0} & \cdots\\
A_{0,-2} & A_{0,-1} & A_{0,0} & \cdots\\
\vdots & \vdots & \vdots & \ddots
\end{pmatrix},
\]
while the regularization matrix $\boldsymbol B$ has the block form
\[
\boldsymbol B
=
\begin{pmatrix}
0 & 0 & \boldsymbol 0\\
0 & 0 & \boldsymbol 0\\
\boldsymbol 0 & \boldsymbol 0 & \boldsymbol B_c
\end{pmatrix},
\qquad
(B_c)_{ij}
=
(\varphi_i,\varphi_j)_F\,,
\quad i,j=0,\ldots,n.
\]
The stationary equations can then be written compactly as
\begin{equation}
(\boldsymbol A+\alpha\boldsymbol B)\,\boldsymbol C
=
\boldsymbol D.
\end{equation}
If no additional boundary conditions are imposed, the regularized solution is
\begin{equation}
\boldsymbol C_\alpha
=
(\boldsymbol A+\alpha\boldsymbol B)^{-1}
\boldsymbol D.
\end{equation}

In practice, the boundary conditions $\rho^c(s_{\rm th})=0$ and $\rho(\Lambda)=\rho^{\mathrm{OPE}}(\Lambda)$ are usually considered. 
Since the hat basis functions satisfy $\varphi_0(x_0)=1$ and $\varphi_n(x_n)=1$, while all other basis functions vanish at the endpoints, these endpoint conditions are equivalent to fixing the corresponding endpoint coefficients,
\begin{equation}
c_0=0,\qquad c_n=\rho^{\mathrm{OPE}}(\Lambda).
\end{equation}
We use the Lagrange multiplier method to incorporate these boundary conditions.
Define $\boldsymbol H$ as a $2\times(n+3)$ matrix,
\begin{equation}
\boldsymbol H
=
\begin{pmatrix}
0&0&1&0&\cdots&0\\
0&0&0&0&\cdots&1
\end{pmatrix},
\end{equation}
whose columns correspond to $c_{-2},c_{-1},c_0,c_1,\ldots,c_n$; the entry $1$ in the first row selects $c_0$, and the entry $1$ in the second row selects $c_n$. We also define
\[
\boldsymbol d
=
\begin{pmatrix}
0\\[1mm]
\rho^{\mathrm{OPE}}(\Lambda)
\end{pmatrix}.
\]
To incorporate the constraints, we introduce the Lagrange multipliers $\boldsymbol\lambda$ and construct the augmented functional
\begin{equation}
\tilde J_\alpha(\boldsymbol C,\boldsymbol\lambda)
=
J_\alpha(\boldsymbol C)
+
\boldsymbol\lambda^{\mathrm T}
\bigl(\boldsymbol H\boldsymbol C-\boldsymbol d\bigr).
\end{equation}
Requiring $\partial\tilde J_\alpha/\partial\boldsymbol C=0$ and $\partial\tilde J_\alpha/\partial\boldsymbol\lambda=0$ yields
\begin{equation}\label{eq:saddle-point}
\begin{pmatrix}
\boldsymbol A+\alpha\boldsymbol B
&
\boldsymbol H^{\mathrm T}
\\
\boldsymbol H
&
\boldsymbol 0
\end{pmatrix}
\begin{pmatrix}
\boldsymbol C_\alpha\\
\boldsymbol\lambda
\end{pmatrix}
=
\begin{pmatrix}
\boldsymbol D\\
\boldsymbol d
\end{pmatrix},
\end{equation}
where $\boldsymbol\lambda$ is the vector of Lagrange multipliers.
Finally, the physical quantities are recovered from the components of $\boldsymbol C_\alpha$ as
\begin{equation}
f_\pi^2
=
m_\pi^2\,c_{-2}\,,
\qquad
f_{a_1}^2
=
m_{a_1}^2\,c_{-1}\,,
\end{equation}
and
\begin{equation}
\rho^c(s)
=
\sum_{i=0}^{n}c_i\,\varphi_i(s).
\end{equation}



\end{document}